\documentclass[prd,twocolumn,showpacs,floatfix,amsmath,nofootinbib,amssymb,floatfix]{revtex4}
\usepackage{graphicx,color,dcolumn,booktabs,bm}
\usepackage{longtable,lscape,booktabs}
\usepackage{multirow,booktabs}
\usepackage{txfonts}
\usepackage{overpic}
\usepackage{amssymb}
\usepackage{indentfirst}
\usepackage{feynmf}   
\usepackage{slashed}  
\usepackage{cases}
\usepackage{color}
\usepackage{multirow}
\usepackage{epstopdf}
\usepackage{xcolor}
\usepackage{soul}
\usepackage{graphicx,color,dcolumn,booktabs,bm}
\usepackage[colorlinks,
            citecolor=blue,
            anchorcolor=red,
            menucolor=red,
            linkcolor=red,
            filecolor=red,
            runcolor=red,
            urlcolor=blue,
            frenchlinks=red]{hyperref}

\begin{document}
\title{Higher states of the $B_c$ meson family}
\author{Ting-Yan Li$^{1,2,3}$}\email{litingyan1213@163.com}
\author{Long Tang$^{4,5,3}$\footnote{ Second Author and First Author contribute equally to this work.}}\email{wangyitl@163.com}
\author{Zheng-Yuan Fang$^{4,3}$}\email{fang1628671420@163.com}
\author{Chao-Hui Wang$^{4,3}$}\email{ch_W187@126.com}
\author{Cheng-Qun Pang$^{4,3,2}$\footnote{Corresponding author}}\email{xuehua45@163.com}


\affiliation{\\$^1$School of Physical Science and Technology, Lanzhou University, Lanzhou 730000, China\\$^2$Lanzhou Center for Theoretical Physics, Key Laboratory of Theoretical Physics of Gansu Province, Lanzhou University, Lanzhou, Gansu 730000, China\\$^3$Joint Research Center for Physics,
Lanzhou University and Qinghai Normal University,
Xining 810000, China\\$^4$College of Physics and Electronic Information Engineering, Qinghai Normal University, Xining 810000, China\\
$^5$Pingwu County Audit Bureau, Pingwu 622550, China}
\begin{abstract}
 In this work, we study higher $B_c$ mesons to the $L=S$, $P$, $D$, $F$, $G$ multiplets using the Cornell potential model, which takes account of the screening effect. The calculated mass spectra of $B_c$ states are in reasonable agreement with the present experimental data. Based on the spectroscopy, partial widths of all allowed radiative transitions and strong decays of each state are also evaluated by applying our numerical wave functions. Comparing our results with the former results, we point out the difference among various models and derive new conclusions obtained in this paper. Our theoretical results are valuable  for searching more $B_c$ mesons in experiments.
\end{abstract}
\date{\today}
\maketitle

\section{Introduction}\label{sec1}
In the past few decades, significant progress has been achieved
in exploring $B_c$ mesons experimentally \cite{ATLAS:2014lga,CMS:2018nak,LHCb:2019hip} and theoretically \cite{1995Heavy,1998Observation}.
In 1998, the ground state of the $B_c$ meson family was first observed by the CDF Collaboration at Fermilab \cite{1998Observation}, with its mass $M_{B_c}=6400\pm390\pm130$ MeV. There was no reported evidence for the excited $B_c$ states until 2014. The ATLS Collaboration reported a structure with the mass of $6842\pm4\pm5$ MeV  \cite{ATLAS:2014lga}, which is regarded as the $B_c^*(2S)$ state in Ref. \cite{ATLAS:2014lga}. However, the mass of this state is lower than the experimental mass of $B_c(2^1S_0)$ state in Particle Data Group (PDG) and the theoretical mass of $B_c(2^1S_0)$ state in other literature, this assignment seems unreasonable. The excited $B_c(2^1S_0)$ state has been observed in the $B^+_c\pi^+\pi^-$ invariant mass spectrum by the LHCb and CMS collaborations and the masses are determined to be $6871.7\pm1.3\pm0.3$ MeV and $6870.6\pm1.4\pm0.3$ MeV in 2019 \cite{CMS:2018nak,LHCb:2019hip}, respectively. So far, different experimental groups have successively observed $S$-wave $B_c$ mesons. The PDG only included two $B_c$ mesons in the summary table \cite{2016Review}. In general, the experimental information about the higher states of $B_c$ meson is still scarce, which is different from the status of charmonium and bottomonium. Exploring the higher states of $B_c$ meson will be constructing the $B_c$ meson family.  

When studying the higher states of charmed/charmed-strange meson, charmonium, and charmed baryon, the role of the coupled-channel effect was found to be important. There are typical examples including $D_{s0}(2317)$ \cite{Luo:2021dvj}, $D_{s1}(2460)$ \cite{2015D(2860)}, $X(3872)$ \cite{Meng:2020cbk, Meng:2021kmi}, and $\Lambda_c(2940)$ \cite{Wang:2020dhf}, where their low mass puzzles can be understood well when the coupled-channel effect is introduced. 
Borrowing the experience mentioned above, we have the reason to believe that the coupled-channel effect cannot be ignored when focusing on the higher states of $B_c$ meson. 

For reflecting the coupled-channel effect generally, we introduce the screening effect \cite{2017A} in the potential model adopted in this work present the mass spectrum for higher states of $B_c$ meson. The detail can be found in the next section. Accompanied with the obtained mass values of these discussed $B_c$ mesons, we can also numerically get the corresponding spatial wave functions for the higher states of $B_c$ meson, which can be applied to the calculation of two-body Okuba–Zweig–Iizuka (OZI) allowed strong decays and radiative decays of them. The information of mass spectrum, strong decays and radiative decays may form a general ideal of the spectroscopy of $B_c$ meson. We hope that the present study can not only reflect the coupled-channel effect in the $B_c$ meson spectroscopy, but also provide theoretical hints to experimental search for higher $B_c$ states in the future.

The structure of this paper is as follows. In {Sec. \ref{sec2}}, we adopt the Cornell potential model including the screening effect to study the mass spectrum of $B_c$ mesons. In {Sec. \ref{sec3}}, we present the
detailed study of the OZI-allowed two-body strong decays and radiative decays of
the discussed $B_c$ mesons.
We analysis decay behaviors of higher excited states of $B_c$ mesons in {Sec. \ref{sec4}}. The paper ends with a conclusion.

\section{mass spectrum}\label{sec2}
 $B_c$ mesons are composed of two heavy quarks.  The heavier mass of the heavy quarks makes that the velocity of the quark in the $B_c$ meson system is relatively small. Thus, we can regard the $B_c$ meson system as a typical non-relativistic system. The mass spectrum of the $B_c$ meson system can be obtained by solving the $\rm{Sch\ddot {o}rdinger}$ equation \cite{1991Bound,Estia1975The}, where the Hamiltonian is
\begin{align}\label{2.1}
H|\psi\rangle=\left(H_{0}+V\right)|\psi\rangle=E|\psi\rangle.
\end{align}
Under the non-relativistic approximation, $H_0$ denotes
\begin{align}
H_{0} = \sum_{i=1}^{2}\left(m_{i}+\frac{p^{2}}{2 m_{i}}\right),
\end{align}
here, $m_{1}$ and $m_{2}$ are the masses of $b$ and $c$ quarks, respectively. We use the Cornell potential as a starting point \cite{E1978Charmonium}, {\it i.e.},
\begin{align}
s(r)=br+c, \\
G(r)=-\frac{4 \alpha_{s}}{3 r},
\end{align}
where $s(r)$ and $G(r)$ are the linear and Coulomb potentials, respectively, and the parameter $c$ denotes the scaling parameter \cite{E1980Charmonium}. When including the screening effect \cite{2017A}, the linear potential becomes
\begin{align}
s(r)^{\prime}=\frac{b\left(1-e^{-\mu r}\right)}{\mu}+c,
\end{align}
where $\mu$ is a screening parameter.

For the spin-dependent term, we refer to it given in the GI model \cite{1985Mesons,1982Potential,1992The,2001QCD,2011Interquark}. The additional screening potential makes that the spin correlation term related to the linear potential also has a corresponding transformation. Thus, we have
\begin{align}
V=H^{conf}+H^{cont}+H^{so}+H^{ten}.\label{2.6}
\end{align}
In Eq. \ref{2.6}, $H^{conf}=s(r)^{\prime}+G(r)$ 
contains the screening potential and Coulomb-like interaction. The colour contact interaction is expressed as
\begin{align}
H^{cont}=\frac{32 \pi \alpha_{s}}{9 m_{1} m_{2}}\left(\frac{\sigma}{\pi^{\frac{1}{2}}}\right)^{3} e^{-\sigma^{2} r^{2}} \boldsymbol{S}_{1} \cdot\boldsymbol{S}_{2}.
\end{align}

\begin{align}
H^{so}=H^{so(cm)}+H^{so(tp)},
\end{align}
which is the spin-orbit interaction, where
\begin{align}
H^{so(cm)}=\frac{4 \alpha_{s}}{3} \frac{1}{r^{3}}\left(\frac{1}{m_{1}}+\frac{1}{m_{2}}\right)^{2} \boldsymbol{L} \cdot \boldsymbol{S}_{1(2)},
\end{align}
\begin{align}
H^{so(tp)}&=-\frac{1}{2 r} \frac{\partial H^{c o n f}}{\partial r}\left(\frac{{\boldsymbol{S}}_{1}}{m_{1}^{2}}+\frac{{\boldsymbol{S}}_{2}}{m_{2}^{2}}\right) \cdot {\boldsymbol{L}}\nonumber\\
&=-\frac{1}{2 r}\left(\frac{4 \alpha_{s}}{3} \frac{1}{r^{2}}+b e^{-\mu r}\right)\left(\frac{1}{m_{1}^{2}}+\frac{1}{m_{2}^{2}}\right) \boldsymbol{L} \cdot \boldsymbol{S}_{1(2)}
\end{align}
is the Thomas precession term with the screening effect. Additionally, we define 
\begin{align}
H^{ten}=\frac{4}{3} \frac{\alpha_{s}}{m_{1} m_{2}} \frac{1}{r^{3}}\left(\frac{3 \boldsymbol{S}_{1} \cdot \boldsymbol{r} \boldsymbol{S}_{2} \cdot \boldsymbol{r}}{\boldsymbol{r}^{2}}-\boldsymbol{S}_{1} \cdot \boldsymbol{S}_{2}\right),
\end{align}
which depicts the colour tensor interaction,
\begin{align}
&\boldsymbol{T}=\frac{ \boldsymbol{S}_{1} \cdot \boldsymbol{r} \boldsymbol{S}_{2} \cdot \boldsymbol{r}}{\boldsymbol{r}^{2}}-\frac{1}{3}\boldsymbol{S}_{1} \cdot \boldsymbol{S}_{2},\\
&\langle T\rangle= \begin{cases}-\frac{L}{6(2 L+3)} & J=L+1 \\ \frac{1}{6} & J=L \\ -\frac{(L+1)}{6(2 L-1)} & J=L-1\end{cases},
\end{align}
where $\boldsymbol{T}$ is the tensor operator,  $\boldsymbol{S}_{1}$ and $\boldsymbol{S}_{2}$ are the spins of the quarks contained by the meson, and $\boldsymbol{L}$ is the orbital angular momentum \cite{2012Interquark}.

The $S$-$L$ coupling causes the mixture of spin singlet and spin triplet states. The Hamiltonian corresponding to the $S$-$L$ term can be sperated as the symmetric and anti-symmetric parts. For the $B_c$ mesons, the results of the symmetric part are zero, we only need to consider the anti-symmetric part 
\begin{align}
H_{\mathrm{anti}}=\frac{1}{4}\left[\left(\frac{4}{3} \frac{\alpha_{s}}{r^{3}}-\frac{b e^{-\mu r}}{r}\right)\left(\frac{1}{m_{1}^{2}}-\frac{1}{m_{2}^{2}}\right) {(\boldsymbol{S}_1-\boldsymbol{S}_2)\cdot{\boldsymbol{L}}}\right]. 
\end{align}
The mixture of states denotes as
\begin{align}
L^{\prime}={ }^{1} L_{J} \cos \theta+{ }^{3} L_{J} \sin \theta, \\
L=-{ }^{1} L_{J} \sin \theta+{ }^{3} L_{J} \cos \theta,
\end{align}
here, $\theta$ is the mixing angle.

The mass spectrum and the wave function of the $B_c$ mesons can be obtained by solving the energy eigenvalue and eigenvector of the  Hamiltonian in Eq. (\ref{2.1}) with the Simple Harmonic Oscillator (SHO) base expanding method.  In configuration and momentum space, SHO wave functions have explicit form respectively

\begin{align}
&\Psi_{n L M_{L}}(\boldsymbol{r})=R_{n L}(r, \beta) Y_{L M_{L}}\left(\Omega_{r}\right), \\
&\Psi_{n L M_{L}}(\boldsymbol{p})=R_{n L}(p, \beta) Y_{L M_{L}}\left(\Omega_{r}\right),
\end{align}
where
\begin{align}
&R_{n L}(r, \beta)=\beta^{\frac{3}{2}} \sqrt{\frac{2 n !}{\Gamma\left(n+L+\frac{3}{2}\right)}}(\beta r)^{L} e^{\frac{-r^{2} \beta^{2}}{2}} L_{n}^{L+\frac{1}{2}}\left(\beta^{2} r^{2}\right), \\
&R_{n L}(p, \beta)=\frac{(-1)^{n}(-i)^{L}}{\beta^{\frac{3}{2}}} e^{-\frac{p^{2}}{2 \beta^{2}}} \sqrt{\frac{2 n !}{\Gamma\left(n+L+\frac{3}{2}\right)}}\left(\frac{p}{\beta}\right)^{L} L_{n}^{L+\frac{1}{2}}\left(\frac{p^{2}}{\beta^{2}}\right),
\end{align}
where $Y_{L M_{L}}\left(\Omega_{r}\right)$ is a spherical harmonic function, $R_{n L}$ ($n=0,1,2,3,\cdots$) is a radial wave function, and $L_{n}^{L+\frac{1}{2}}(x)$ denotes a Laguerre polynomial.

We considered the landau full in our calculation. To avert the condition, we gave a cutoff distance $r_c$ for solving the $\rm{Sch\ddot {o}rdinger}$ equation.  $r_c$ is determined by fitting the spectrum and the value of it is listed in Table \ref{tab1}. The detail for this $r_c$ are introduced in Ref \cite{Deng:2016ktl}.

In order to obtain the mass spectra of $B_c$ mesons, we should fix the correspondent parameters involved in the adopted potential model by fitting the experimental data. Here, we select the reported $B_c$ mesons, charmonia, bottomonia, and charmed/charmed-stranged and bottom/bottom-strange mesons in the experiment \cite{2016Review}, which are listed in Table \ref{experimental}. 
By performing the $\chi^{2}$ fitting defined as 
\begin{align}
\chi^{2}=\sum _i \frac{(\text {Th$_i$}-\text {Exp$_i$})^2}{\text {Error$_i$}^2},
\end{align}
finally these input parameters can be obtained\footnote{In the $\chi^2$ fitting, we should find the minimum value of $\chi^{2}$, where these parameters can fixed. In this work, $\chi^{2}/d.o.f.$ is 5656, where $d.o.f.$ is the degree of freedom which is 24. In Table \ref{Parameters}, the obtained parameters in the potential model are given.}, where Th, Exp, and Error represent the theoretical, experimental data, and fitting error, respectively.

\renewcommand{\arraystretch}{1.1}
\begin{table}[htbp]
\caption{The experimental mass values and fitted results of heavy flavor mesons in this work (unit: MeV). The experimental data of $B_c$ mesons are derived from PDG \cite{2016Review}, and the other mesons are also derived from PDG \cite{2016Review}. {\color{black}Here, Exp and Th represent the  experimental data and theoretical data, respectively.}  }
\label{experimental}
\[\begin{array}{cccccc}
\toprule[1pt]\toprule[1pt]
\text{Meson} & \text{State}   &\text{Exp}  & \text{Th} \\
\midrule[1pt]
B_c    & 1^1S_0 & 6274.47\pm0.27\pm0.17 &\color{black}{6269}\\
       & 2^1S_0  & 6871.2\pm1.0 &\color{black}{6886} \\
\hline
c\bar{c}    & 1^1S_0 & 2983.9\pm0.5 &\color{black}{3007} \\
            & 2^1S_0 & 3637.6\pm1.2&\color{black}{3645} \\
            & 1^3S_1 & 3096.9\pm0.006 &\color{black}{3100}\\
            & 2^3S_1 & 3686.097\pm0.01 &\color{black}{3686}\\
            & 1^1P_1 & 3525.38\pm0.11 &\color{black}{3522}\\
            & 1^3P_0 & 3414.71\pm0.3 &\color{black}{3406}\\
            & 1^3P_1 & 3510.67\pm0.05&\color{black}{3515}\\
            & 1^3P_2 & 3556.17\pm0.07 &\color{black}{3540}\\
\hline
b\bar{b}    & 1^1S_0 & 9399\pm2.3 &\color{black}{9372}\\
            & 2^1S_0 & 9999\pm3.5 &\color{black}{10025}\\
            & 1^3S_1 & 9460.3\pm0.26 &\color{black}{9414}\\
            & 2^3S_1 & 10023.2\pm0.31 &\color{black}{10037}\\
            & 1^1P_1 & 9899.3\pm0.8 &\color{black}{9933}\\
            & 1^3P_0 & 9859.4\pm0.42\pm0.31 &\color{black}{9888}\\
            & 1^3P_1 & 9892.8\pm0.26\pm0.31 &\color{black}{9928}\\
            & 1^3P_2 & 9912.2\pm0.26\pm0.31 &\color{black}{9950}\\
\hline
D           & 1^1S_0 & 1864.84\pm0.5 &\color{black}{1889}\\
            & 2^1S_0 & 2564\pm20 &\color{black}{2583}\\
            & 1^3S_1 & 2006.85\pm0.05&\color{black}{2007}\\
            & 1^3P_2 & 2460.7\pm0.4 &\color{black}{2438}\\
\hline
D_s         & 1^1S_0 & 1969\pm1.4 &\color{black}{1995}\\
            & 1^3S_1 & 2112.2\pm0.4 &\color{black}{2111}\\
            & 2^3S_1 & 2708.3\pm4 &\color{black}{2736}\\
            & 1^3P_1 & 2459.6\pm0.9 &\color{black}{2538}\\
            & 1^3P_2 & 2569.1\pm0.8 &\color{black}{2544}\\
            & 1^3D_1 & 2859\pm12\pm24 &\color{black}{2866}\\
            & 1^3D_3 & 2860.5\pm26\pm6.5 &\color{black}{2828}\\
\hline
B           & 1^1S_0 & 5279.25\pm0.26&\color{black}{5272}\\
            & 1^3S_1 & 5324.7\pm0.21 &\color{black}{5319}\\
            & 1^3P_2 & 5737.2\pm0.7 &\color{black}{5719}\\
\hline
B_s         & 1^1S_0 & 5366.84\pm0.14 &\color{black}{5275}\\
            & 1^3S_1 & 5415.8\pm1.5 &\color{black}{5321}\\
            & 1^3P_2 & 5839.92\pm0.14 &\color{black}{5751}\\
\bottomrule[1pt]\bottomrule[1pt]
\end{array}\]
\end{table}

\renewcommand{\arraystretch}{1.2}
\begin{table}[htbp]\footnotesize
	\centering
\caption{The fitted parameters in the potential model adopted in this work.}
\label{Parameters}

\[\begin{array}{ccccccc}
\toprule[1pt]\toprule[1pt]
\text{Parameter} & \text{Value} & \text{Parameter} & \text{Value}  &\\
\midrule[1pt]
m_b    &  \color{black}{5.368}~\text{GeV}   & m_u,\,m_d & \color{black}{0.606}~\text{GeV}  &\\
m_c    &  \color{black}{1.984}~\text{GeV}  & m_s    &  \color{black}{0.780}~\text{GeV}  & \\
\alpha_s & \color{black}{0.3930}  & \sigma & \color{black}{1.842}~\text{GeV}  &  \\
b       & \color{black}{0.2312}~\text{GeV}^2  & c   & \color{black}{-1.1711}~\text{GeV}  &  \\
\mu & \color{black}{0.0690}~\text{GeV}  &\color{black}{ r_c }&\color{black}{0.3599 ~\text{GeV}^{-1} }& \\
\bottomrule[1pt]\bottomrule[1pt]
\end{array}\]\label{tab1}
\end{table}

Using the parameters in Table \ref{Parameters}, we obtain the mass spectra of $B_c$ mesons as shown in Table \ref{mass} and give the masses of these $S$-wave, $P$-wave, $D$-wave, $F$-wave, and $G$-wave states. The calculated $S$-, $P$-, $D$-, $F$-, and $G$-wave states are considered as $n+L\le6$ (where $n$ is the radial quantum number and $n=1$ is the ground state). In addition, the mixing angles for some mixtures of states are also given in Table \ref{mass}. Of course, we notice that there were former theoretical studies of the $B_c$ mass spectrum. Thus, we 
also make the comparison of our results with 
other theoretical results from Refs.  \cite{Li:2019tbn,Zeng:1994vj,Soni:2017wvy,Monteiro:2016ijw,1994Mesons,Ebert:2002pp,2004Spectroscopy,Gershtein:1994dxw}.
\par

It is found that the masses of the low-lying $1S$-, $1P$-, $1D$-, and $1F$-wave $B_c$ states predicted in this work are compatible with the other potential model predictions.
For the higher mass states, such as $5S$- and $4P$-wave states,
the masses predicted by us are smaller than those predicted with the non-relativistic quark model in Ref.~\cite{Li:2019tbn}.
The strong screening effect for the large angular momentum and larger  average distance between quark pair  leads to the mass for the higher exited states in this work  lower than other potential models. 
\par

The states with quantum  numbers $^{1}S_0$ and $^{3}S_1$ are the partner of $B_c$ mesons. The ground state $B_c(1S)$ and radial excited state $B_c(2S)$ have been established in the experiment, and their measured average masses are $6274.47\pm 0.27\pm 0.17$ MeV and $6871.2\pm1.0$ MeV \cite{Workman:2022ynf}, respectively, which are in  agreement with our theoretical values in Table \ref{experimental}.
We find that our predict  mass for the $1^3S_1$ state is 6322 MeV, which matches the result of Lattice QCD of 6331$\pm$10 MeV \cite{Mathur:2018epb}. For the $2^3S_1$ state, we give its mass 6907 MeV. The  mass for the $3^1S_0$ is about 7216 MeV.
The hyperfine mass splitting between the spin-singlet and spin-triplet states $\Delta m (nS)=m[B_c(n^3S_1)]-m[B_c(n^1S_0)]$, which is caused by the spin-dependent interaction and often be used to test various potential models. 
For the $1S$ state,  $\Delta m(1S)\simeq 60 \ \ \mathrm{MeV}$, which is very close to $55$ MeV predicted in Ref. \cite{Li:2019tbn}. 
The predicted masses for the other higher $S$-wave states compared with other works are also given in Table \ref{mass}. 
\par
 For the $P$-wave states of $B_c$mesons, we should consider the mixture for the $n^1P_1$ and $n^3P_1$ states, which have the same radial quantum number $n$. By including the mixture of the $n^1P_1$ and $n^3P_1$ states, the physical states $P^{\prime}_1$ and $P_1$ can be formed. The masses of $1P^{\prime}_1$ and $1P_1$ states are 6761 and  6770 MeV, respectively, where the mixing angle  $\theta_{1p}= -24.3^{\circ}$  is obtained. The masses of $1P^{\prime}_1$ and $1P_1$ states are very close to those predicted values in Ref. \cite{Li:2019tbn}.
The masses of $B_c(2P)$ and $B_c(2P^{\prime})$ state  are also consistent with the other predictions with potential models~\cite{Li:2019tbn,Zeng:1994vj,Monteiro:2016ijw}. 
The predicted masses of $3P$  and $4P$ states are  smaller approximately 100 $-$ 200 MeV than the  predictions in Ref. \cite{Monteiro:2016ijw}, which is resulted in the screening effect. The masses of  $5P$  are also calculated.

Except  $B_c(4D)$, the absolute mixing angle of $nD$ and  $nF$ states  consist with Ref. \cite{Li:2019tbn}.
In addition, the $G$-wave mass of the $B_c$ mesons also be calculated, there are few work about the mass of $B_c(1G)$ and  $B_c(2G)$ states. The masses of $1G$ and  $1G^\prime$  states are located at a range of 7427$-$7443 MeV according to our estimates. 
\par
We hope that this work can help for the research of higher states of $B_c$ mesons in the future. 
After providing the information of mass spectrum for the discussed higher states of the $B_c$ meson family, we should further study their decay behaviors as illustrated in the following section.

\begin{longtable*}[tbp]
{p{0.9cm}p{1.3cm}p{1.4cm}p{1.5cm}p{1.5cm}p{1.5cm}p{1.5cm}p{1.5cm}p{1.5cm}p{1.5cm}p{1.5cm}}
\caption{Predicted masses (MeV) of $B_c$ states compared with other model predictions and data.
The mixing angles between $B_{c}(n^{3}L_{J})$ and $B_{c}(n^{1}L_{J})$ obtained in this work are also presented.\label{mass}}\\
\toprule
 \toprule
 \hline
 \hline
State~~~~&Ours~~~~ &LZ~\cite{Li:2019tbn}~~~~ & ZVR~\cite{Zeng:1994vj}~~~~& SJSCP~\cite{Soni:2017wvy} &MBV~\cite{Monteiro:2016ijw}~~~~&EQ~\cite{1994Mesons}~~~~
&EFG~\cite{Ebert:2002pp}~~~~& GI~\cite{2004Spectroscopy}~~~~& KLT~\cite{Gershtein:1994dxw}\\
 \midrule
  \midrule
  \hline
 \endfirsthead {\bf continue~\ref{mass}}\\
 \toprule
  \toprule
  \hline
   \hline
State~~~~&Ours~~~~ &LZ~\cite{Li:2019tbn}~~~~ & ZVR~\cite{Zeng:1994vj}~~~~& SJSCP~\cite{Soni:2017wvy} &MBV~\cite{Monteiro:2016ijw}~~~~&EQ~\cite{1994Mesons}~~~~
&EFG~\cite{Ebert:2002pp}~~~~& GI~\cite{2004Spectroscopy}~~~~& KLT~\cite{Gershtein:1994dxw}\\
   \midrule
      \midrule
 \hline
  \endhead \endfoot
  \bottomrule
    \bottomrule
             \hline
             \hline
     \endlastfoot
$3 ^1S_{0}$~~~~&\color{black}{7261}~~~~&7239        ~~~~~&7240    ~~~~&7306    ~~~~ &7308     ~~~~&7244    ~~~~&7193    ~~~~&7250    ~~~~&$\cdots$\\
$4 ^1S_{0}$~~~~&\color{black}{7551}~~~~&7540        ~~~~~&7550    ~~~~&7684    ~~~~ &7713     ~~~~&7562    ~~~~&$\cdots$~~~~&$\cdots$~~~~&$\cdots$\\
$5 ^1S_{0}$~~~~&\color{black}{7790}~~~~&7805        ~~~~~&$\cdots$~~~~&8025    ~~~~ &8097     ~~~~&$\cdots$~~~~&$\cdots$~~~~&$\cdots$~~~~&$\cdots$\\
$6 ^1S_{0}$~~~~&\color{black}{7994}~~~~&8046        ~~~~~&$\cdots$~~~~&8340    ~~~~ &8469     ~~~~&$\cdots$~~~~&$\cdots$~~~~&$\cdots$~~~~&$\cdots$\\
$1 ^3S_{1}$~~~~&\color{black}{6322}~~~~&6326~~~~~&6340    ~~~~&6321    ~~~~ &6357     ~~~~&6337    ~~~~&6332    ~~~~&6338    ~~~~&6317    \\
$2 ^3S_{1}$~~~~&\color{black}{6907}~~~~&6890        ~~~~~&6900    ~~~~&6900    ~~~~ &6897     ~~~~&6899    ~~~~&6881    ~~~~&6887    ~~~~&6902  \\
$3 ^3S_{1}$~~~~&\color{black}{7275}~~~~&7252        ~~~~~&7280    ~~~~&7338    ~~~~ &7333     ~~~~&7280    ~~~~&7235    ~~~~&7272    ~~~~&$\cdots$\\
$4 ^3S_{1}$~~~~&\color{black}{7561}~~~~&7550        ~~~~~&7580    ~~~~&7714    ~~~~ &7734     ~~~~&7594    ~~~~&$\cdots$~~~~&$\cdots$\\
$5 ^3S_{1}$~~~~&\color{black}{7798}~~~~&7813        ~~~~~&$\cdots$~~~~&8054    ~~~~ &8115     ~~~~&$\cdots$~~~~&$\cdots$~~~~&$\cdots$~~~~&$\cdots$\\
$6 ^3S_{1}$~~~~&\color{black}{8001}~~~~&8054        ~~~~~&$\cdots$~~~~&8368    ~~~~ &8484     ~~~~&$\cdots$~~~~&$\cdots$~~~~&$\cdots$~~~~&$\cdots$\\
$1P'_1$~~~~&\color{black}{6761}~~~~&6776        ~~~~~&6740    ~~~~&$\cdots$~~~~ &6734     ~~~~&6736    ~~~~&6749    ~~~~&6750    ~~~~&6729 \\
$1P_1$ ~~~~&\color{black}{6770} ~~~~&6757        ~~~~~&6730    ~~~~&$\cdots$~~~~ &6686     ~~~~&6730    ~~~~&6734    ~~~~&6741    ~~~~&6717 \\
$\theta_{1p}$&\color{black}{-24.3}$^{\circ}$&35.5$^{\circ}$&$\cdots$~~~~&$\cdots$~~~~&$\cdots$~~~~&$\cdots$~~~~&20.4$^{\circ}$~~~~&22.4$^{\circ}$~~~~&$\cdots$\\
$2P'_1$ ~~~~&\color{black}{7156}~~~~&7150        ~~~~~&7150    ~~~~&$\cdots$~~~~ &7173     ~~~~&7142    ~~~~&7145    ~~~~&7150    ~~~~&7124 \\
$2P_1$ ~~~~&\color{black}{7164}~~~~&7134        ~~~~~&7140    ~~~~&$\cdots$~~~~ &7137     ~~~~&7135    ~~~~&7126    ~~~~&7145    ~~~~&7113 \\
$\theta_{2p}$&\color{black}{-28.4}$^{\circ}$&38.0$^{\circ}$&$\cdots$~~~~&$\cdots$~~~~&$\cdots$~~~~&$\cdots$~~~~&{23.2}$^{\circ}$~~~~&18.9$^{\circ}$~~~~&$\cdots$\\
$3P'_1$ ~~~~&\color{black}{7458} ~~~~&7458        ~~~~~&7470    ~~~~&$\cdots$~~~~ &7572     ~~~~&$\cdots$~~~~&$\cdots$~~~~&$\cdots$~~~~&$\cdots$\\
$3P_1$ ~~~~&\color{black}{7466}    ~~~~&7441        ~~~~~&7460    ~~~~&$\cdots$~~~~ &7546     ~~~~&$\cdots$~~~~&$\cdots$~~~~&$\cdots$~~~~&$\cdots$\\
$\theta_{3p}$&\color{black}{-30.2}$^{\circ}$&39.7$^{\circ}$&$\cdots$~~~~&$\cdots$~~~~&$\cdots$~~~~&$\cdots$~~~~&$\cdots$~~~~&$\cdots$~~~~&$\cdots$\\
$4P'_1$ ~~~~&\color{black}{7708 } ~~~~&7727        ~~~~~&7740    ~~~~&$\cdots$~~~~ &7942     ~~~~&$\cdots$~~~~&$\cdots$~~~~&$\cdots$~~~~&$\cdots$\\
$4P_1$ ~~~~&\color{black}{7715}   ~~~~&7710        ~~~~~&7740    ~~~~&$\cdots$~~~~ &7943     ~~~~&$\cdots$~~~~&$\cdots$~~~~&$\cdots$~~~~&$\cdots$\\
$\theta_{4p}$&\color{black}{-31.0}$^{\circ}$&39.7$^{\circ}$&$\cdots$~~~~&$\cdots$~~~~&$\cdots$~~~~&$\cdots$~~~~&$\cdots$~~~~&$\cdots$~~~~&$\cdots$\\
$5P'_1$ ~~~~&\color{black}{7921 } ~~~~&$\cdots$       ~~~~&$\cdots$    ~~~~&$\cdots$~~~~&$\cdots$     ~~~~&$\cdots$~~~~&$\cdots$~~~~&$\cdots$~~~~&$\cdots$\\
$5P_1$ ~~~~&\color{black}{7927}   ~~~~&$\cdots$       ~~~~&$\cdots$    ~~~~&$\cdots$~~~~&$\cdots$     ~~~~&$\cdots$~~~~&$\cdots$~~~~&$\cdots$~~~~&$\cdots$\\
$\theta_{5p}$&\color{black}{-31.6}$^{\circ}$&$\cdots$~~~~&$\cdots$~~~~&$\cdots$~~~~&$\cdots$~~~~&$\cdots$~~~~&$\cdots$~~~~&$\cdots$~~~~&$\cdots$\\
$1 ^3P_{0}$~~~~&\color{black}{6712}~~~~&6714        ~~~~~&6680    ~~~~&6686    ~~~~ &6638     ~~~~&6700    ~~~~&6699    ~~~~&6706    ~~~~&6683 \\
$1 ^3P_{2}$~~~~&\color{black}{6783}~~~~&6787        ~~~~~&6760    ~~~~&6712    ~~~~ &6737     ~~~~&6747    ~~~~&6762    ~~~~&6768    ~~~~&6743 \\
$2 ^3P_{0}$~~~~&\color{black}{7118}~~~~&7107        ~~~~~&7100    ~~~~&7146    ~~~~ &7084     ~~~~&7108    ~~~~&7091    ~~~~&7122    ~~~~&7088\\
$2 ^3P_{2}$~~~~&\color{black}{7175}~~~~&7160        ~~~~~&7160    ~~~~&7173    ~~~~ &7175     ~~~~&7153    ~~~~&7156    ~~~~&7164    ~~~~&7134 \\
$3 ^3P_{0}$~~~~&\color{black}{7427}~~~~&7420        ~~~~~&7430    ~~~~&7536    ~~~~ &7492     ~~~~&$\cdots$~~~~&$\cdots$~~~~&$\cdots$~~~~&$\cdots$\\
$3 ^3P_{2}$~~~~&\color{black}{7476}~~~~&7464        ~~~~~&7480    ~~~~&7565    ~~~~ &7575     ~~~~&$\cdots$~~~~&$\cdots$~~~~&$\cdots$~~~~&$\cdots$\\
$4 ^3P_{0}$~~~~&\color{black}{7682}~~~~&7693        ~~~~~&7710    ~~~~&7885    ~~~~ &7970     ~~~~&$\cdots$~~~~&$\cdots$~~~~&$\cdots$~~~~&$\cdots$\\
$4 ^3P_{2}$~~~~&\color{black}{7724}~~~~&7732        ~~~~~&7760    ~~~~&7915    ~~~~ &7970     ~~~~&$\cdots$~~~~&$\cdots$~~~~&$\cdots$~~~~&$\cdots$\\
$5 ^3P_{0}$~~~~&\color{black}{7899}~~~~&$\cdots$        ~~~~&$\cdots$    ~~~~&$\cdots$    ~~~~&$\cdots$     ~~~~&$\cdots$~~~~&$\cdots$~~~~&$\cdots$~~~~&$\cdots$\\
$5 ^3P_{2}$~~~~&\color{black}{7936}~~~~&$\cdots$       ~~~~&$\cdots$   ~~~~&$\cdots$    ~~~~&$\cdots$     ~~~~&$\cdots$~~~~&$\cdots$~~~~&$\cdots$~~~~&$\cdots$\\
$1D'_2$  ~~~~&\color{black}{7046} ~~~~&7032        ~~~~~&7030    ~~~~&$\cdots$~~~~ &7003     ~~~~&7012    ~~~~&7079    ~~~~&7036    ~~~~&7124  \\
$1D_2$  ~~~~&\color{black}{7037}   ~~~~&7024        ~~~~~&7020    ~~~~&$\cdots$~~~~ &6974     ~~~~&7009    ~~~~&7077    ~~~~&7041    ~~~~&7113 \\
$\theta_{1D}$&\color{black}{-41.7}$^{\circ}$&45.0$^{\circ}$~~~~&$\cdots$~~~~&$\cdots$~~~~&$\cdots$~~~~&-35.9$^{\circ}$~~~~&$\cdots$~~~~&44.5$^{\circ}$~~~~&$\cdots$\\
$2D'_2$  ~~~~&\color{black}{7365}  ~~~~&7347        ~~~~~&7360    ~~~~&$\cdots$~~~~ &7408     ~~~~&$\cdots$~~~~&$\cdots$~~~~&$\cdots$~~~~&$\cdots$\\
$2D_2$  ~~~~&\color{black}{7360}   ~~~~&7343        ~~~~~&7360    ~~~~&$\cdots$~~~~ &7385     ~~~~&$\cdots$~~~~&$\cdots$~~~~&$\cdots$~~~~&$\cdots$\\
$\theta_{2D}$&\color{black}{-42.6}$^{\circ}$&45.0$^{\circ}$~~~~&$\cdots$~~~~&$\cdots$~~~~&$\cdots$~~~~&$\cdots$~~~~&$\cdots$~~~~&$\cdots$~~~~&$\cdots$\\
$3D'_2$  ~~~~&\color{black}{7627}  ~~~~&7623        ~~~~~&7650    ~~~~&$\cdots$~~~~ &7783     ~~~~&$\cdots$~~~~&$\cdots$~~~~&$\cdots$~~~~&$\cdots$\\
$3D_2$  ~~~~&\color{black}{77623 } ~~~~&7620        ~~~~~&7650    ~~~~&$\cdots$~~~~ &7781     ~~~~&$\cdots$~~~~&$\cdots$~~~~&$\cdots$~~~~&$\cdots$\\
$\theta_{3D}$&\color{black}{43.6}$^{\circ}$&45.0$^{\circ}$~~~~&$\cdots$~~~~&$\cdots$~~~~&$\cdots$~~~~&$\cdots$~~~~&$\cdots$~~~~&$\cdots$~~~~&$\cdots$\\
$4D'_2$ ~~~~&\color{black}{7849 }  ~~~~&$\cdots$       ~~~~~&$\cdots$    ~~~~&$\cdots$~~~~ &$\cdots$     ~~~~&$\cdots$~~~~&$\cdots$~~~~&$\cdots$~~~~&$\cdots$\\
$4D_2$  ~~~~&\color{black}{7846 } ~~~~&$\cdots$        ~~~~~&$\cdots$    ~~~~&$\cdots$~~~~ &$\cdots$     ~~~~&$\cdots$~~~~&$\cdots$~~~~&$\cdots$~~~~&$\cdots$\\
$\theta_{4D}$&\color{black}{-44.6}$^{\circ}$&$\cdots$~~~~&$\cdots$~~~~&$\cdots$~~~~&$\cdots$~~~~&$\cdots$~~~~&$\cdots$~~~~&$\cdots$~~~~&$\cdots$\\
$1 ^3D_{1}$~~~~&\color{black}{7037}~~~~&7020        ~~~~~&7010    ~~~~&6998    ~~~~ &6973     ~~~~&7012    ~~~~&7072    ~~~~&7025    ~~~~&7088 \\
$1 ^3D_{3}$~~~~&\color{black}{7042}~~~~&7030        ~~~~~&7040    ~~~~&6990    ~~~~ &7004     ~~~~&7005    ~~~~&7081    ~~~~&7045    ~~~~&7134 \\
$2 ^3D_{1}$~~~~&\color{black}{7357}~~~~&7336        ~~~~~&7350    ~~~~&7403    ~~~~ &7377     ~~~~&$\cdots$~~~~&$\cdots$~~~~&$\cdots$~~~~&$\cdots$\\
$2 ^3D_{3}$~~~~&\color{black}{7364}~~~~&7348        ~~~~~&7370    ~~~~&7399    ~~~~ &7410     ~~~~&$\cdots$~~~~&$\cdots$~~~~&$\cdots$~~~~&$\cdots$\\
$3 ^3D_{1}$~~~~&\color{black}{7619}~~~~&7611        ~~~~~&7640    ~~~~&7762    ~~~~ &7761     ~~~~&$\cdots$~~~~&$\cdots$~~~~&$\cdots$~~~~&$\cdots$\\
$3 ^3D_{3}$~~~~&\color{black}{7627}~~~~&7625        ~~~~~&7660    ~~~~&7761    ~~~~ &7796     ~~~~&$\cdots$~~~~&$\cdots$~~~~&$\cdots$~~~~&$\cdots$\\
$4 ^3D_{1}$~~~~&\color{black}{7842}~~~~&$\cdots$    ~~~~~&$\cdots$    ~~~~&$\cdots$    ~~~~ &$\cdots$     ~~~~&$\cdots$~~~~&$\cdots$~~~~&$\cdots$~~~~&$\cdots$\\
$4 ^3D_{3}$~~~~&\color{black}{7850}~~~~&$\cdots$   ~~~~~&$\cdots$    ~~~~&$\cdots$    ~~~~ &$\cdots$     ~~~~&$\cdots$~~~~&$\cdots$~~~~&$\cdots$~~~~&$\cdots$\\
$1F'_3$ ~~~~&\color{black}{7261 }~~~~&7240        ~~~~~&7250    ~~~~&$\cdots$~~~~ &$\cdots$ ~~~~&$\cdots$~~~~&$\cdots$~~~~&7266    ~~~~&$\cdots$\\
$1F_3$  ~~~~&\color{black}{7248}  ~~~~&7224        ~~~~~&7240    ~~~~&$\cdots$~~~~ &$\cdots$ ~~~~&$\cdots$~~~~&$\cdots$~~~~&7276    ~~~~&$\cdots$\\
$\theta_{1F}$&\color{black}{-41.8}$^{\circ}$&41.4$^{\circ}$~~~~&$\cdots$~~~~&$\cdots$~~~~&$\cdots$~~~~&$\cdots$~~~~&$\cdots$~~~~&\color{black}{41.4}$^{\circ}$~~~~&$\cdots$\\
$2F'_3$ ~~~~&\color{black}{7537}   ~~~~&7525        ~~~~~&7550    ~~~~&$\cdots$~~~~ &$\cdots$ ~~~~&$\cdots$~~~~&$\cdots$~~~~&7571    ~~~~&$\cdots$\\
$2F_3$ ~~~~&\color{black}{7526}   ~~~~&7508        ~~~~~&7540    ~~~~&$\cdots$~~~~ &$\cdots$ ~~~~&$\cdots$~~~~&$\cdots$~~~~&7563    ~~~~&$\cdots$\\
$\theta_{2F}$&\color{black}{-41.2}$^{\circ}$&43.4$^{\circ}$~~~~&$\cdots$~~~~&$\cdots$~~~~&$\cdots$~~~~&$\cdots$~~~~&$\cdots$~~~~&$\cdots$~~~~&$\cdots$\\
$3F'_3$~~~~&\color{black}{7770}   ~~~~&7779        ~~~~~&7810    ~~~~&$\cdots$~~~~ &$\cdots$ ~~~~&$\cdots$~~~~&$\cdots$~~~~&$\cdots$~~~~&$\cdots$\\
$3F_3$ ~~~~&\color{black}{7762}    ~~~~&7768        ~~~~~&7800    ~~~~&$\cdots$~~~~ &$\cdots$ ~~~~&$\cdots$~~~~&$\cdots$~~~~&$\cdots$~~~~&$\cdots$\\
$\theta_{3F}$&\color{black}{-41.3}$^{\circ}$&42.4$^{\circ}$~~~~&$\cdots$~~~~&$\cdots$~~~~&$\cdots$~~~~&$\cdots$~~~~&$\cdots$~~~~&$\cdots$~~~~&$\cdots$\\
$1 ^3F_{2}$~~~~&\color{black}{7258}~~~~&7235        ~~~~~&7240    ~~~~&7234    ~~~~ &$\cdots$ ~~~~&$\cdots$~~~~&$\cdots$~~~~&7269    ~~~~&$\cdots$\\
$1 ^3F_{4}$~~~~&\color{black}{7249}~~~~&7227        ~~~~~&7250    ~~~~&7244    ~~~~ &$\cdots$ ~~~~&$\cdots$~~~~&$\cdots$~~~~&7271    ~~~~&$\cdots$\\
$2 ^3F_{2}$~~~~&\color{black}{7533}~~~~&7518        ~~~~~&7540    ~~~~&7607    ~~~~ &$\cdots$ ~~~~&$\cdots$~~~~&$\cdots$~~~~&7565    ~~~~&$\cdots$\\
$2 ^3F_{4}$~~~~&\color{black}{7528}~~~~&7514        ~~~~~&7550    ~~~~&7617    ~~~~ &$\cdots$ ~~~~&$\cdots$~~~~&$\cdots$~~~~&7568    ~~~~&$\cdots$\\
$3 ^3F_{2}$~~~~&\color{black}{7767}~~~~&7730        ~~~~~&7800    ~~~~&7946    ~~~~ &$\cdots$ ~~~~&$\cdots$~~~~&$\cdots$~~~~&$\cdots$~~~~&$\cdots$\\
$3 ^3F_{4}$~~~~&\color{black}{7764}~~~~&7771        ~~~~~&7810    ~~~~&7956    ~~~~ &$\cdots$ ~~~~&$\cdots$~~~~&$\cdots$~~~~&$\cdots$~~~~&$\cdots$\\
$1G'_4$ ~~~~&\color{black}{7443}  ~~~~&$\cdots$        ~~~~&$\cdots$    ~~~~&$\cdots$~~~~&$\cdots$     ~~~~&$\cdots$~~~~&$\cdots$~~~~&$\cdots$~~~~&$\cdots$\\
$1G_4$ ~~~~&\color{black}{7427}   ~~~~&$\cdots$        ~~~~&$\cdots$    ~~~~&$\cdots$~~~~&$\cdots$     ~~~~&$\cdots$~~~~&$\cdots$~~~~&$\cdots$~~~~&$\cdots$\\
$\theta_{1G}$&\color{black}{-41.5}$^{\circ}$&$\cdots$~~~~&$\cdots$~~~~&$\cdots$~~~~&$\cdots$~~~~&$\cdots$~~~~&$\cdots$~~~~&$\cdots$~~~~&$\cdots$\\
$2G'_4$ ~~~~&\color{black}{7687}  ~~~~&$\cdots$        ~~~~&$\cdots$    ~~~~&$\cdots$~~~~&$\cdots$     ~~~~&$\cdots$~~~~&$\cdots$~~~~&$\cdots$~~~~&$\cdots$\\
$2G_4$ ~~~~&\color{black}{7675}   ~~~~&$\cdots$        ~~~~&$\cdots$    ~~~~&$\cdots$~~~~&$\cdots$    ~~~~&$\cdots$~~~~&$\cdots$~~~~&$\cdots$~~~~&$\cdots$\\
$\theta_{2G}$&\color{black}{-41.5}$^{\circ}$&$\cdots$~~~~&$\cdots$~~~~&$\cdots$~~~~&$\cdots$~~~~&$\cdots$~~~~&$\cdots$~~~~&$\cdots$~~~~&$\cdots$\\
$1 ^3G_{3}$~~~~&\color{black}{7441}~~~~&$\cdots$       ~~~~~&$\cdots$    ~~~~&$\cdots$    ~~~~ &$\cdots$ ~~~~&$\cdots$~~~~&$\cdots$~~~~&$\cdots$~~~~&$\cdots$\\
$1 ^3G_{5}$~~~~&\color{black}{7428}~~~~&$\cdots$        ~~~~~&$\cdots$    ~~~~&$\cdots$    ~~~~ &$\cdots$ ~~~~&$\cdots$~~~~&$\cdots$~~~~&$\cdots$~~~~&$\cdots$\\
$2 ^3G_{3}$~~~~&\color{black}{7686}~~~~&$\cdots$        ~~~~~&$\cdots$    ~~~~&$\cdots$    ~~~~ &$\cdots$ ~~~~&$\cdots$~~~~&$\cdots$~~~~&$\cdots$~~~~&$\cdots$\\
$2 ^3G_{5}$~~~~&\color{black}{7675}~~~~&$\cdots$       ~~~~~&$\cdots$    ~~~~&$\cdots$    ~~~~ &$\cdots$ ~~~~&$\cdots$~~~~&$\cdots$~~~~&$\cdots$~~~~&$\cdots$\\
\end{longtable*}

\section{Theoretical models of decay behaviors}\label{sec3}
Next, we give the necessary formulas for calculating two-body OZI-allowed strong decays and radiative transitions.

\subsection{Two-body OZI-allowed strong decays}
The quark pair creation (QPC) model was first proposed by Micu and has been widely used to calculate the strong two-body decay allowed by OZI rule after further development \cite{1969Decay,2005Radiative,Blundell:1996as,Yu:2011ta,2011Categorizing,2012Non}.
For a decay process $A\to B+C$, the transition matrix is defined by \cite{2013Towards,2012Mass,2015Light,2016Strong,2019Study}
\begin{equation}\label{3.6}
\langle BC|\mathcal{T}|A \rangle = \delta ^3(\boldsymbol{{P}_B}+\boldsymbol{{P}_C)}\mathcal{M}^{{M}_{J_{A}}M_{J_{B}}M_{J_{C}}},
\end{equation}

the amplitude
$\mathcal{M}^{M_{J_A}M_{J_B} M_{J_C}}(\mathbf{P})$ can be derived. $\mathcal{T}$ can be expressed as
\begin{align}\label{3.7}
\mathcal{T}& = -3\gamma \sum_{m}\langle 1m;1~-m|00\rangle\int d {\boldsymbol{p}}_3d {\boldsymbol{p}}_4\delta ^3 ({\boldsymbol{p}}_3+{\boldsymbol{p}}_4) \nonumber \\
 & ~
 \quad\times \mathcal{Y}_{1m}\left(\frac{{\boldsymbol{p}}_3-{\boldsymbol{p}}_4}{2}\right)\chi _{1,-m}^{34}\phi _{0}^{34}
\left(\omega_{0}^{34}\right)_{ij}b_{3i}^{\dag}({\boldsymbol{p}}_3)d_{4j}^{\dag}({\boldsymbol{p}}_4),
\end{align}
which is the transition operator and it can describe the creation of a quark-antiquark pair from vacuum.
Where $b_3^{\dag}(d_{4}^{\dag})$ is quark (antiquark) creation operator. $\chi$, $\phi$ and $\omega$ denote the spin, flavour, and color wave functions, respectively. $\gamma$ is a dimensionless constant depicting the generation rate of a quark-antiquark pair from a vacuum.  $\mathcal{Y}_{\ell m}({p})={|{p}|^{\ell}}Y_{\ell
m}({p})$ is a solid spherical harmonic function. Using the Jacobi-Wick formula, the $\mathcal{M}^{M_{J_A}M_{J_B} M_{J_C}}(\mathbf{P})$ is converted into the partial wave amplitudes $\mathcal{M}^{JL}$, and it can be expressed as
\begin{align}\label{3.8}\begin{split}
\mathcal{M}^{J L}({\boldsymbol{P}})=& \frac{\sqrt{4 \pi(2 L+1)}}{2 J_{A}+1} \sum_{M_{J_{B}} M_{J_{C}}}\left\langle L 0 ; J M_{J_{A}} \mid J_{A} M_{J_{A}}\right\rangle \\
& \times\left\langle J_{B} M_{J_{B}} ; J_{C} M_{J_{C}} \mid J_{A} M_{J_{A}}\right\rangle \mathcal{M}^{M_{J_{A}} M_{J_{B}} M_{J_{C}}}.
\end{split}\end{align}
Then the strong decay partial width for a given decay mode of $A \rightarrow B+C$ reads as
\begin{align}\label{3.9}
\Gamma=\frac{\pi}{4} \frac{|P_E|}{m_{A}^{2}} \sum_{J, L}\left|\mathcal{M}^{J L}({\boldsymbol{P}})\right|^{2},
\end{align}
where $\mathrm{m}_{A}$ is the mass of the initial meson $A$. 

In addition, the meson wave function is defined as a mock state, {\it i.e.},
\begin{align}\label{3.10}\begin{split}
\left|A\left(n^{2 S+1} L_{J M_{J}}\right)\left({\boldsymbol{p}}_{A}\right)\right\rangle=& \sqrt{2 E} \sum_{M_{S}, M_{L}}\left\langle L M_{L} S M_{S} \mid J M_{J}\right\rangle \chi_{S M_{S}}^{A} \\
& \times \phi^{A} \omega^{A} \int d {\boldsymbol{p}}_{1} d {\boldsymbol{p}}_{2} \delta^{3}\left({\boldsymbol{p}}_{A}-{\boldsymbol{p}}_{1}-{\boldsymbol{p}}_{2}\right) \\
& \times \Psi_{n L M_{L}}^{A}\left({\boldsymbol{p}}_{1}, {\boldsymbol{p}}_{2}\right)\left|q_{1}\left({\boldsymbol{p}}_{1}\right) \bar{q}_{2}\left({\boldsymbol{p}}_{2}\right)\right\rangle,
\end{split}
\end{align}
where the spatial wave function $\psi_{n L M_{L}}({p})$ of the meson is obtained by solving Eq. (\ref{2.1}).

 We can filter out  OZI-allowed two-body strong decay channels of these discussed $B_c$ mesons. For these strong decays with a pair of strange quarks from the vacuum, the strength of strange pair creation has the relation $\gamma_s=\gamma/\sqrt{3}$, where $\gamma=6.947$ is taken from Ref. \cite{Li:2019tbn}. The decay widths of the two-body strong decay are calculated and the results are collected in section \ref{sec4}. The corresponding branching ratios of the decay widths of these discussed $B_c$ mesons are also given.

\subsection{Radiative transitions}

The partial width for the $E1$ transitions of $B_c$ mesons in the non-relativistic quark model is given by Ref. \cite{2004Spectroscopy}
\begin{align}
\Gamma(i \rightarrow f+\gamma)=\frac{4}{3}\left\langle e_{Q}\right\rangle^{2} \alpha \omega^{3} C_{f i} \delta_{S S^{\prime}}|\langle f|r| i\rangle|^{2},
\end{align}
where
\begin{align}
C_{f i}=\operatorname{Max}\left(L, L^{\prime}\right)\left(2 J^{\prime}+1\right)\left\{\begin{array}{ccc}
L^{\prime} & J^{\prime} & S \\
J & L & 1
\end{array}\right\}^{2},
\end{align}
\begin{align}
\left\langle e_{Q}\right\rangle=\frac{m_{b} e_{c}-m_{c} e_{\bar{b}}}{m_{b}+m_{c}},
\end{align}
where the quantum numbers with prime and without prime denote final state and initial state, respectively. $m_c$ and $m_b$ are the masses of the quarks. The charge of the quark is $e_b=-1/3$, $e_c= 2/3$. $\omega$ is the energy of the photon. The following equation can be obtained by the conservation of energy and momentum
\begin{align}
M_{i}=\sqrt{M_{f}^{2}+\omega^{2}}+\omega.
\end{align}
Here the radial wave function $R_{nL}(r)$ is same with one used in the strong decay.

\section{analysis of excited states of $B_c$ mesons}\label{sec4}
\begin{table}[htbp]
\caption{{\color{black}Partial widths and branching ratios of OZI-allowed strong decay for $3^1S_0$ $-$ $6^1S_0$ states of $B_c$ mesons. The width results are in units of MeV.}}
\vspace{-12pt}
\label{s0}
{\color{black}
\[\begin{array}{clrrrr}
\toprule[1pt]\toprule[1 pt]
\text{State}  & \text{Channels}   &\text{This work}   & \text{ Br ($\%$)}  &$\cite{Li:2019tbn}$  &\text{ Br ($\%$)}\\
\midrule[0.7pt]
 3^1S_0       & B^*D                    & 173     & 100             & 161           & 100 \\
              &\text{Total}             & 173     & 100             & 161           & 100\\
 4^1S_0       & B^{*}D^{*}              & 66      & 50.3            & 104           & 54\\
              & BD^*                    & 40.8    & 31.1            & 34.9          & 18.3\\
              & B^*D                    & 18.4    & 14              & 0.14         & 0.1\\
              & B_{s}^{*}D_{s}^*        & 4.01    & 3.06            &15.5           &8.1\\
              & B_{s}D_{s}^*            & 1.28    & 0.976           & 5.8           & 3.1\\
              & B_{s}^{*}D_{s}          & 0.64    & 0.49            & 6.7            & 3.5\\
              & BD_0^*(2300)            & \cdots  & \cdots          & 24            & 12.6\\
              &\text{Total}             &131      & 100               & 191            &100 \\
5^1S_0         & B^*D^{'}_{1}(2420) & {\color{black}47.3 }   & {\color{black}24.4}     & 70.9          & 17.2 \\
             & BD^*_{2}(2460)     & 42.5    & {\color{black}21.9}            & 48.2          & 11.7 \\
             & B^*_{2}(5747)D^*   & 39.9    & {\color{black}20.5}& 56.5          & 13.7 \\
             & B^*D^*             & 31.9    & {\color{black}16.4}           & 2.28          & 0.6 \\
             & BD^*               & 10.5    &{\color{black}5.46}             & 1.5           & 0.4 \\
             & B^*_{2}(5747)D     & {\color{black}8.98}    & {\color{black}4.63}            & 27.6          & 6.7 \\
             
              & {\color{black}B_{1}(5721)D^{*}}         & {\color{black}5.88}   & \cdots           & 6.2           & 1.5 \\
             
             & B^*D_{1}(2430)     &{\color{black}3.99}    & {\color{black}2.06}             & 12.3          & 3.0 \\
            
             & {\color{black}B^*D  }             & {\color{black}0.789}   & {\color{black}0.406}            & 24.5          & 5.9 \\
             
             & B^*D_{2}^*(2460)   & {\color{black}0.707}   & {\color{black}0.364}            & 25.7          & 6.2 \\
            
             & BD^*_{0}(2300)     & 0.584   &{\color{black}0.3}            & 23.5          & 5.7 \\
            & B_{s}D^*_{s}       & {\color{black}0.218}    & {\color{black}0.112}           & 4.65          & 1.1 \\
             
             & B(1^3P_0)D         & \cdots  & \cdots           & 18.6          & 4.5 \\
             & B^{*}_{s}D^{*}_s   & \cdots  & \cdots           & 5.75          & 1.4 \\
             & \text{Total}       & {\color{black}194}     & 100              & 413           & 100\\
6^1S_0       & B^*_{2}(5747)D        & 15.8    & 30.4               & 4.85     & 1.3 \\
             & B^*D(2550)            & 10.1    & 19.4               &\cdots    &\cdots\\
             & B^*D^*                &{\color{black} 9.02 }   &{\color{black} 17.4 }                & 24.3     & 6.7 \\
             & B^*D^*_{2}(2460)      & {\color{black}6.78}    &{\color{black} 13.1 }              & 23.5     & 6.5 \\
             
             & {\color{black} B^*D_1(2430)}          & {\color{black} 1.59}    &{\color{black}  3.07  }             & 41.4     & 11.4 \\
             
              &  {\color{black}BD^* }                 &  {\color{black}1.53 }   &  {\color{black}2.95}               & 24.3     & 6.7 \\
             
             & B^*D^{'}_{1}(2420)    & {\color{black}1.37}   &  {\color{black}2.64   }            & 46.8     & 13 \\
               &{\color{black} B^{*}_{2}(5747)D^*}    &{\color{black} 0.968}   & {\color{black}1.87 }               & 20.6     &5.7 \\
             & B_{1}(5721)D^*        & {\color{black}0.831  } &{\color{black} 1.6}             & 24.7     & 6.8 \\
             
             & B^*_{s}D_{s1}(2536)   & {\color{black}0.624}   & {\color{black}1.2}               &4.14    &1.1\\
              & B^*D                  &  {\color{black}0.617 }  &  {\color{black}1.19}               & 44.4     & 12 \\
             & B_{s1}(5830)D^*_{s}   &  {\color{black}0.541}   & 1.46               &0.03    &0.01\\
          
             & B_{s}D^*_{s0}(2317)   & 0.429   & 0.825              & 6.62     & 1.8 \\
             & BD^*_{0}(2300)        & 0.35    & 0.673              & 13.2     & 3.6 \\
             & BD^*_{2}(2460)        & 0.148   & 0.285              & 28.9     & 8 \\
                      & B(1P^{'})D^{*}        &\cdots   &\cdots              & 28.3     & 7.8 \\
             & B(1^3P_0)D            &\cdots   &\cdots              & 11.3     & 3.1 \\ 
             &   \text{Total}        & 52      &  100               &361       &100\\
 \bottomrule[1pt] \bottomrule[1pt]
\end{array}\]
}
\end{table}

In this section, we will discuss the radiative transition (here, we neglect the smaller M1 transitions and only consider E1 transitions) and the OZI-allowed two-body strong decay of $B_c$ mesons.In the final states of  strong decay, we  consider $1P$ or $1P'$ mixed of the resonances $D_{1}(2430)$, $D_{1}(2420)$, $D_{s1}(2536)$, $B_{1}(5721)$ and $B_{s1}(5830)$  \cite{Chen:2016spr, Chen:2022asf}.  {\color{black}In our calculation of the strong decays, we neglect the value of Br less than one percent.}

\subsection{$S$-wave states} The details of S-wave decay for the $B_c$ mesons are given in Tables \ref{s0}-\ref{6s1} and Table \ref{dianfusheS}. For completeness we give a brief overview of these decays. For the $B_c(1S)$ and  $B_c(2S)$ states below the $BD$ threshold, we consider the $E1$ transitions of the $B_c(2S)$. In our calculation,
$\Gamma_{B_c(2^1S_0)\to 1P_1 \gamma}\simeq 1\ \ \mathrm{keV}$, $\Gamma_{B_c(2^1S_0)\to 1P_1^\prime \gamma}\simeq 8\ \ \mathrm{keV}$. The $E1$ transitions of the $S$-wave states $B_c^*(2S)$ are also discussed. The $E1$ transitions of $S$-wave states of $B_c$ mesons are also discussed in other literature for establish these missing $B_c$ states in PDG. Our predicted partial width
$\Gamma_{B_c(2^3S_1)\to {\color{black}1P_1} \gamma}\simeq {\color{black}4.2}\ \ \mathrm{keV}$, 
our result is very close to {\color{black}4.7} keV predicted in Ref. \cite{2004Spectroscopy}. 
The total radiative decay width of $B_c(2^3S_1)$ is about 16 keV.

Comparing the calculation of the strong decay $B_c(3^1S_0)\to B^*D$ predicted in~\cite{Li:2019tbn} with this work, the value of the decay width is estimated to be $\Gamma_{total}\sim 173$ MeV, which is larger than the result 161 MeV in Ref. \cite{Li:2019tbn}. This process is the important finial state  to determine the $B_c(3^1S_0)$ state in future experiments. The two-body decay information of $4^{1}S_0$, $5^{1}S_0$ and $6^{1}S_0$ state can also be obtained in Table \ref{s0}. 
 Ratios of $BD^*$, $B^*D^*$, and $B^*D$ for the $4^{1}S_0$ state are 0.31, 0.5, and 0.14, respectively. {\color{black}$B^*D^{\prime}_1(2420)$} is the dominant decay mode of $5^{1}S_0$, and  $BD_2^*(2460)$ and $B_2^*(5747)D^*$ are its important  final  state. For the  $6^{1}S_0$ state, the dominant decay channel is  $B_2^*(5747)D$. The total decay width of $4^{1}S_0$, $5^{1}S_0$, and $6^{1}S_0$ are 131, {\color{black}194}, and 52 MeV, respectively.
The $B^*D$ and $BD$ channels are the dominant decay modes of $3^{3}S_1$. The total width of $3^{3}S_1$  state is 142 MeV, larger than the predicted result of $\sim$ 133 MeV in Ref. \cite{Li:2019tbn}.
$B^*D^*$ is the  largest  decay mode of $4^{3}S_1$. $B^*D$ and $BD^*$ have important contribution to the total width of $4^{3}S_1$ state.   $5^{3}S_1$ has a broad width of $\sim$ 189 MeV.
Similar to the  $6^{1}S_0$ state, $6^{3}S_1$  has the dominant decay channel $B^* D_2^*(2460)$. Other details for the higher $S$-wave states $B_c(n^1S_0)$ and $B_c(n^3S_1)$ are given in Table \ref{s0}, Table\ref{s1} and Table \ref{6s1}, respectively.

\begin{table}
\caption{{\color{black}Partial widths and branching ratios of OZI-allowed strong decay for $3^3S_1$ $-$ $5^3S_1$ states of $B_c$ mesons. The width results are in units of MeV.}}
\vspace{-12pt}
\label{s1}
{\color{black}\[\begin{array}{clrrrr}
\toprule[1pt]\toprule[1pt]
\text{State} & \text{Channels} & \text{This work} &\text{ Br ($\%$)} &$ \cite{Li:2019tbn} $&\text{ Br ($\%$)}\\
\midrule[0.7pt]
 3^3S_1 & B^*D                    & 108   & 75.8    & 105     & 79  \\
        & BD                      & 34.4  & 24.2    & 28      & 21 \\
        &\text{Total}             & 142   &100      & 133     & 100 \\
 4^3S_1 & B^*D^*                  & 85.1  & 67.6    & 112     & 66 \\        
        & BD^*                    & 25.5  & 20.2    & 17.0    & 10 \\
        & B^*D                    & {\color{black}8.23}  & {\color{black}6.56 }   & 0.41    & 0.2 \\
        & B^*_{s}D^*_{s}          & 5.08  & 4.03    & 26.9    & 16  \\
        & BD                      &{\color{black}0.673} & {\color{black}0.537}   &4.53     &2.7  \\
        & B^*_{s}D_s              & {\color{black}0.614}   & 0.499   &5.29     &3.1 \\
        
        & B_{s}D_{s}              & 0.574 & 0.456   &2.81     &1.6  \\
        & B_{s}D^*_{s}            &{\color{black}0.488}   & 0.38    &1.83     &1.1  \\
        &  \text{Total}           & 126   & 100     & 171     &100 \\
 5^3S_1 & B^*_{2}(5747)D^*        & 64.3  & {\color{black}34.1}&{\color{black}96}    &{\color{black}24}\\
  & B^*D^*          & {\color{black}32.7}    & {\color{black}17.3}   &0.19      &0.05 \\
         & B^*D_{1}(2430)          & {\color{black}20.6}    & {\color{black}10.9 }  &6.8      &1.7 \\
        & BD^*_{2}(2460)          & 20.5  & {\color{black}10.9 }&17.34    &4.3 \\
                & B^*D_{1}^{\prime}(2420) & {\color{black}12.3}  & {\color{black}6.52}  &32.9     &8.2\\
        & B^*D^*_{2}(2460)        & 11    & {\color{black}5.77}    &69.87    &17\\
        & B^*_{2}(5747)D          &{\color{black}6.69}  & {\color{black}3.54}        &16.1     &4\\
        & BD^*                    &{\color{black}5.38} & {\color{black}2.85}    &2.65     &0.7\\
        & B_{1}(5721)D^*          & {\color{black}3.78}  & 1.9     &5.19     &1.3 \\
        & B_1(5721)D              &  {\color{black}3.77} & 1.9   &13.34    &3.3\\
             & BD_{1}(2430)            & {\color{black}3.25}  &  {\color{black}1.72}    &0.63     &0.2 \\
        & BD_{1}^{\prime}(2420)   & {\color{black}2.62} &  {\color{black}1.39}   &0.89     &0.2\\
        & BD                      & {\color{black}0.491} &  {\color{black}0.26} &15.81    &3.9\\
        & B^*D_{0}^*(2300)        & 0.408 &  {\color{black}0.215}&18.32    &4.6 \\
        
                & B(1P^{'})D^*            & \cdots& \cdots  &53.93    &13.4 \\
        & B(1P')D                 & \cdots& \cdots  &18.96    &4.7 \\
                & B^{*}D                  & \cdots& \cdots  &20.18    &5 \\
        &  \text{Total}           &{\color{black}189}   & 100     & 401     &100 \\     
 \bottomrule[1pt] \bottomrule[1pt]
\end{array}\]}
\end{table}

\begin{table}
\caption{{\color{black}Partial widths and branching ratios of OZI-allowed strong decay for $6^3S_1$ state of $B_c$ mesons. The width results are in units of MeV.}}
\vspace{-12pt}
\label{6s1}
{\color{black}\[\begin{array}{clrrrr}
\toprule[1pt]\toprule[1pt]
\text{State} & \text{Channels} & \text{This work} & \text{ Br ($\%$)} &$ \cite{Li:2019tbn}$&\text{ Br ($\%$)}\\
\midrule[0.7pt]
     6^3S_1   
          & B^*D^*                 &{\color{black} 7.65}  &{\color{black} 16.2}                  & 37.9     & 10.2\\
     & B_{2}^*(5747)D         & {\color{black}7.56}  & {\color{black}16 }                 & 1.09     & 0.3\\
     
        & B^*D(2550)             & {\color{black}5.45}  &{\color{black} 11.6}                  &\cdots    &\cdots \\
        & B^*D^*_{2}(2460)       & {\color{black}5.03}   & {\color{black}10.7}                  & 39.9     & 10.7\\
          & B_{1}(5721)D           & {\color{black}4.56}  & {\color{black}9.67}     & 11.1     & 3\\
        & BD^{'}_{1}(2420)       & {\color{black}3.98}  &{\color{black} 8.44}                  & 9.37     & 2.5\\
        & BD_{1}(2430)           & {\color{black}3.76}  & {\color{black}7.97}                  & 16.6     & 4.4\\
        & BD                     & {\color{black}1.85} & {\color{black}3.92}                  & 17.6     & 4.7\\
        
                & B^*D_{1}(2430)         &  {\color{black}1.58} &  {\color{black}3.35}                 & 27.9     & 7.5\\

        & B^*D^{'}_{1}(2420)     &{\color{black} 1.05} & {\color{black}2.23}                 & 30.5     & 8.2\\
        & B^*D                   & {\color{black}1.01}  & {\color{black}2.14}                 & 31       & 8.3\\
         & B_{1}(5721)D^*         &  {\color{black}0.757}  &  {\color{black}1.61}               & 17       & 4.6\\
          & BD^*                   & {\color{black}0.477}& {\color{black}1.01}               & 19.1     & 5.1\\
        & B^*_{s}D^*_{s}         & {\color{black}0.473} & {\color{black}1}                & 2.96     & 0.8\\
        
         & B^*D^*_{0}(2300)       & 0.428 & {\color{black}0.91}                & 12.9     & 3.5\\
        & BD^*_{2}(2460)         & {\color{black}0.389} & {\color{black}0.825}                & 14.1     & 3.8\\
       
        & B^*_{s}D_{s1}(2536)    & \cdots & \cdots                & 3.06   & 0.8\\

                & B_{2}^*(5747)D^{*}     &\cdots &\cdots                 & 34       & 9.1 \\
        & B(1P')D^{*}            & \cdots&\cdots                 & 21.2     & 5.7\\
        & B(1^{3}P_{0})D^{*}     & \cdots& \cdots                & 10.9     & 3\\
        & \text{Total}           & 48    & 100                   & 372      & 100 \\
 \bottomrule[1pt] \bottomrule[1pt]
\end{array}\]}
\end{table}

\subsection{$P$-wave states}
The decay behaviors of $B_c(nP)$ with $n=1,2,3,4,5$ are given in Tables \ref{p1},\ref{p4} and \ref{dianfusheS}. 
The $1P$-wave of $B_c$ states  have no OZI-alowed two body strong decay channel and 
the mainly decay models are radiative decays. We have calculated the partial decay widths for the $E1$ transitions $1P\to 1S\gamma$, $2P\to 1S\gamma$, $2P\to 2S\gamma$, and $2P\to 1D\gamma$. 
Combine with the processes ${B_c(2^1S_0)\to 1P_1 \gamma}$ and ${B_c(2^1S_0)\to 1P_1^\prime \gamma}$, 
${B_c(2^1S_0)\to B_c\gamma\gamma}$ and $B_c\gamma$ final channel for $B_c(1P_1)$ and $B_c(1P^\prime_1)$  maybe helpful to search  the $B_c(1P_1)$ and $B_c(1P^\prime_1)$, i.e., the $B_c(1P_1)$ and $B_c(1P_1')$ states might be first found in the $B_c\gamma$ and $B_c\gamma\gamma$  final states via their radiative transitions. 
In Table \ref{dianfusheS}, we can see that the $B_c(1^3P_0)$ and $B_c(1^3P_2)$ states dominantly decay into $B_c^*\gamma$ final state with  decay widths 60 keV and 90 keV, respectively. The approximate  width  of the $B_c(2P_1)\to 2^1S_0\gamma$, $B_c(2P_1)\to 2^3S_1\gamma$ and $B_c(2^3P_2)\to {\color{black}1^3D_3}\gamma$ decay processes are {\color{black}14} keV, 38 keV, and 10 keV respectively. 
  $B_c(2^3P_0)$ dominantly decays into $B_c^*(2S)\gamma$ final state.
The details of E1 transitions  for $B_c(nP)$ states can be found in  Table \ref{dianfusheS}.

 As shown in Table \ref{p4}, $B_c(2^3P_2)$ only has one two body strong decay mode $BD$, its width is about 4 MeV, which may be helpful for experiment to search this $B_c$ state. For the $3P^{\prime}_1$ and $3P_1$ states, $B^*D^*$ and and $BD^*$  are all their  main  decay  modes, which have the total width of {\color{black}201} MeV and {\color{black}265} MeV. The values of the total width for these two states are {\color{black} larger than} the results in Ref. \cite{Li:2019tbn} and the decay channels are similar. Compared to the results in Ref.\cite{Li:2019tbn}, the total widths  for the $4^3P_2$, $4P^{\prime}_1$ and $4P_1$ are smaller. But the the total width  for the $4^3P_0$ is larger and our total width is of {\color{black}84.9} MeV. We also calculate the decay width for the $5P-$wave $B_c$ states.

 Except $B_c(2^3P_2)$, $B_c(5P_1^\prime)$, $B_c(5P_1)$ and $B_c(5^3P_2)$, $B^*D^*$ final channel is the dominant decay mode for the other open-charmed  decays of $B_c(nP)$ such as $B_c(3P_1^\prime)$, $B_c(3P_1)$, $B_c(4P_1^\prime)$, $B_c(4P_1)$, $B_c(3^3P_0)$, $B_c(3^3P_2)$, $B_c(4^3P_0)$, $B_c(4^3P_2)$ and $B_c(5^3P_0)$.
$B^*D^*$ mode will play an important role in the discovery for the higher $B_c(nP)$ states.
The other OZI allowed strong decay channels can be found in Table \ref{p1} and \ref{p4}. 

\begin{table}
\renewcommand\arraystretch{1.1}
\caption{{\color{black}Partial widths and branching ratios of OZI-allowed strong decay for the mixture of the $3^1P_1$ and $3^3P_1$ states $-$ $5^1P_1$ and $5^3P_1$ states. The width results are in units of MeV.}}
\vspace{-12pt}
\label{p1}
{\color{black}\[\begin{array}{clrrrr}
\toprule[1pt]\toprule[1pt]
\text{State} & \text{Channels} & \text{ This work} & \text{ Br ($\%$)} & $\cite{Li:2019tbn}$ & \text{ Br ($\%$)}\\
\midrule[0.7pt]

 3P^{'}_{1}    & B^*D^* & {\color{black}134} & {\color{black}66.8}           &129   &69.4\\
 \text{} & BD^* & {\color{black}62.1} & {\color{black}30.9}    &32    &17.2\\ 
 \text{} & B^*_{s}D_s & {\color{black}2.86}& {\color{black}1.43}    &11.1  &6     \\
 \text{} & B^*D & {\color{black}1.66}&{\color{black}1.827}     &13.6  &7.3\\
 \text{} & \text{Total} & {\color{black}201} & 100     &185   &100 \\
 
 3 \text{P}_{1} & B^*D^* &{\color{black} 164} & {\color{black}61.8}     &145    &65.8\\
 \text{} & BD^* & 55.8 & 21            &62     &28.1\\
 \text{} & B^*D & 45.2 & 17            &9.3    &4.3\\ 
 \text{} & B^*_{s}D_s & 0.227 & 0.09   &4.0    &1.8\\
 \text{} & \text{Total} & 265 & 100      &220    &100\\
 
 4 P^{'}_{1} &B^{*}_2(5747)D  &39.9  &43.8   &6.55 &4.6  \\    
 &B^{*}D^{*}  &{\color{black}29.1} &{\color{black}32}   &6.55 &4.6  \\        
 \text{} & BD^* & 16.6 & 32.9        &11.9  &8.4\\
 \text{} & B^*D & {\color{black}2.98} & {\color{black}3.27}   &41.6  &29.1\\
 
  
 \text{} & B_sD^*_{s} & {\color{black}0.552} & {\color{black}0.607}      & 6.2  &4.3\\

 \text{} & B^*_{s}D^*_{s} &  {\color{black}0.211} &  {\color{black}0.232} &9.09  &6.3\\

& B(1^3P_{2})D    &\cdots &\cdots    &36.6  &25.6\\
 & BD^*_{0}(2300)   &\cdots &\cdots    &13.6  &9.5\\
  & B(1^3P_{0})D  &\cdots &\cdots      &10.4  &7.3\\ 
       & B_{s}D^*_{s0}(2317) &\cdots &\cdots  &4.75  &3.3\\
\text{} & \text{Total} & {\color{black}91} & 100    &143   &100\\

 4 \text{P}_{1} & B^*D^* & {\color{black}21.9} & {\color{black}54.4}   &0.86 &0.7 \\
 &BD_{1}(2430)      &{\color{black}7.28} &{\color{black}18.1 }      &23.03  &18.3\\ 
 \text{} & BD^* &{\color{black}4.63} &{\color{black}11.5 }     &3.7    &2.9\\ 
 \text{} & B^*_{s}D_s &{\color{black}1.79} &{\color{black}4.45 }    &4.4    &3.5\\ 
 \text{} & B^*D &{\color{black}1.51} &{\color{black}3.75 } &24.5   &19.4\\ 
 \text{} & B_sD^*_{s} & {\color{black} 0.63} &  {\color{black}1.57}   &6.78   &5.4\\ 
 \text{} & B^*_{s}D^*_{s} &  {\color{black}0.366} &  {\color{black}0.909} &6.66   &5.3\\
  &BD_{1}^{\prime}(2420)     &\cdots &\cdots    &15.32  &12.2\\ 
 &B^*_2(5747)D & {\color{black}0.314}&{\color{black}0.78} &15 &11.9\\
&B(1^3P_0)D^{*}  &\cdots &\cdots   &11.8   &9.4\\ 
&B^{*}D_0^*(2300)   &\cdots &\cdots  &10.02  &8.0\\
  &B_{1}(5721)D      &\cdots &\cdots       &3.32   &2.6\\
   \text{} & \text{Total} & {\color{black}40.2} & 100    &126  &100\\

 5 \text{P}'_{1} & B^*D(2550) & {\color{black}24.1} & {\color{black}30.2}  &\cdots &\cdots\\
 \text{} & {\color{black}B^*_2(5747)D} & {\color{black}14.3} & {\color{black}17.9}  &\cdots &\cdots\\
  \text{} & {\color{black}B^*D^*_2(2460)} & {\color{black}8.5} & {\color{black}10.6}  &\cdots &\cdots\\
 \text{} & B^*D^* & {\color{black}6.61} & {\color{black}8.27}  &\cdots &\cdots\\
  \text{} & {\color{black}B^*D} & {\color{black}5.51} & {\color{black}6.9} &\cdots &\cdots\\
   \text{} & {\color{black}B^*_2(5747)D} & {\color{black}5.39} & {\color{black}6.75} &\cdots &\cdots\\ 
 \text{} & BD^* & {\color{black}4.25} & {\color{black}5.32} &\cdots &\cdots\\
  \text{} & {\color{black}BD^*_2(2460)} & {\color{black}3.43} & {\color{black}4.29} &\cdots &\cdots\\
  \text{} & {\color{black}B^*D^{\prime}_1(2420)} & {\color{black}3.21} & {\color{black}4.02} &\cdots &\cdots\\
   \text{} & {\color{black}B_1(5721)D} & {\color{black}1.37} & {\color{black}1.71} &\cdots &\cdots\\
 \text{} & \text{Total} & {\color{black}79.9} & 100 &\cdots &\cdots \\
 
 5 \text{P}_{1} &{\color{black}B^*_2(5747)D}& {\color{black}16.7} & {\color{black}26.3}    &\cdots &\cdots\\
 &{\color{black}B^*D^*_2(2460)}& {\color{black}12.7} & {\color{black}20}    &\cdots &\cdots\\
 & B^*D(2550) & {\color{black}10.8} &  {\color{black}17 }   &\cdots &\cdots\\
 \text{} & B^*D^* & {\color{black}5.87} & {\color{black}9.25}             &\cdots &\cdots\\
 &{\color{black}B_1(5721)D}& {\color{black}3.68} & {\color{black}5.8}    &\cdots &\cdots\\
 &{\color{black}B^*D^{\prime}_1(2420)}& {\color{black}2.78} & {\color{black}4.38}    &\cdots &\cdots\\
 &{\color{black}BD^*_2(2460)}& {\color{black}2.69} & {\color{black}4.24}    &\cdots &\cdots\\
 \text{} & BD^* & {\color{black}1.52} & {\color{black}2.39}     &\cdots &\cdots\\
  &{\color{black}BD_{1}(2430) }     &{\color{black}0.926} &{\color{black}1.46 }       &\cdots &\cdots\\
  \text{} & {\color{black}B^*_sD_{s}} & {\color{black} 0.756} &  {\color{black}1.19}   &\cdots &\cdots\\

 \text{} & \text{Total} & {\color{black}63.5} & 100  &\cdots &\cdots\\
 \bottomrule[1pt] \bottomrule[1pt]
\end{array}\]}
\end{table}

\begin{table}
\renewcommand\arraystretch{1.1}
\caption{{\color{black}Partial widths and branching ratios of OZI-allowed strong decay for $2^3P_2-5^3P_2$-wave $B_c$ states. The width results are in units of MeV.}}
\vspace{-12pt}
\label{p4}
{\color{black}\[\begin{array}{clrrrr}
\toprule[1pt]\toprule[1pt]
\text{State}&\text{Channels}&\text{ This work}&\text{ Br ($\%$)} &$\cite{Li:2019tbn}$&\text{ Br ($\%$)}\\
\midrule[0.7pt]
 2^3 \text{P}_2 & BD& 4.01 & 100  &\cdots &\cdots\\
 \text{} & \text{Total} & 4.01 & 100      &\cdots &\cdots\\
 3^3 \text{P}_0 & B^*D^* & 211 & 78.2   &255   &93\\
 \text{} & BD & {\color{black} 58.4} &  {\color{black}21.6}        &9.6   &3.5\\
 \text{} & B_sD_s &  {\color{black}0.892} &  {\color{black}0.33 }          &9.7   &3.5  \\
 \text{} & \text{Total} & 270 & 100  &274   &100\\
 
 3^3 \text{P}_2 & B^*D^* & 135 & 84.4    &146    &74\\
 \text{} & BD^* &  {\color{black}17.2} &  {\color{black}10.7}           &3.4    &1.7\\ 
 \text{} & BD &  {\color{black}4.76} &  {\color{black}2.97}      &22     &11.1\\
 \text{} & B^*_sD_s & 1.76 & 1.1   &7.8    &4\\
 \text{} & B_sD_s &  {\color{black}0.942} &  {\color{black}0.588}      &2.7    &1.4\\ 
 \text{} & B^*D &  {\color{black}0.437} & {\color{black} 0.273 }       &16     &8.1\\
 \text{} & \text{Total} & 160 & 100    &198    &100\\
 
 4^3 \text{P}_0 & B^*D^* &  {\color{black}73.5} & {\color{black} 86.6}      &14     &26.4\\ 
 \text{} & BD &  {\color{black}8.61} &  {\color{black}10.1}       &13.6   &25.6\\
 \text{} & B_sD_s & {\color{black} 2.77} &  {\color{black}3.26 }&7.16   &13.5\\
 &B(1P^{'})D    &\cdots &\cdots    &7.66   &14.4\\
    &B^{*}D_{0}^*(2300) &\cdots &\cdots &5.5    &10.4\\
   &B^{*}_{s}D^{*}_{s}&\cdots &\cdots &4.6    &8.7\\
 \text{} & \text{Total} &  {\color{black}84.9} & 100 &53     &100\\
 
 4^3 \text{P}_2 & B^*D^* & {\color{black}33.4} & {\color{black}42} &7.82   &4.1\\
&{\color{black} B^*_2(5747)D} & {\color{black}17.3} & {\color{black}21.7} &{\color{black}20.2}   &{\color{black}10.6}\\
 &BD_{1}^{\prime}(2420)   &{\color{black}12.4} &{\color{black}15.6}    &13.1   &6.9\\
 \text{} & BD & {\color{black}6.59} & {\color{black}8.28}  &21.76  &11.4\\
 \text{} & B^*D & {\color{black}4.36}& {\color{black}5.48}    &30.1   &15.8\\
  &B_{1}(5721)D     &{\color{black}3.27}&{\color{black}4.11}      &6.95   &3.6\\
    &BD_{1}(2430)      &{\color{black}0.787} &{\color{black}0.989}       &6.61   &3.5\\
 \text{} & B^*_sD^*_s &  {\color{black}0.501} & {\color{black} 0.629}  &11.1   &5.8\\
 \text{} & B_sD^*_s & 0.4 & {\color{black}0.498}       &2.34   &1.2\\
 &B(1P^{'})D    &\cdots &\cdots    &27.7   &14.5\\
  &B^{*}D_{0}^*(2300)  &\cdots &\cdots&10.1   &5.3\\ 
  &B^{*}D_{1}(2430)   &\cdots &\cdots      &9.22   &4.8\\ 
  &B(1^3P_{0})D^{*}  &\cdots &\cdots &8.8    &4.6\\
 \text{} & \text{Total} & {\color{black}79.6} & 100          &190    &100\\
 
  5^3 \text{P}_0 & {\color{black}B^*D^*_2(2460)} & {\color{black}32.1} & {\color{black}35.5}  & \cdots& \cdots\\ 
 \text{} & {\color{black}B^*D^*} & {\color{black}22.5} & {\color{black}24.9}   & \cdots& \cdots\\
  \text{} & {\color{black}BD(2550)} & {\color{black}16.4} & {\color{black}18.1}   & \cdots& \cdots\\
   \text{} & {\color{black}B^*_2(5747)D^*} & {\color{black}9.98} & {\color{black}11}   & \cdots& \cdots\\
    \text{} & {\color{black}B^*D_1(2430)} & {\color{black}3.55} & {\color{black}3.93}   & \cdots& \cdots\\
     \text{} & {\color{black}B_1(5721)D^*} & {\color{black}1.93} & {\color{black}2.13}   & \cdots& \cdots\\
 \text{} & B_sD_s & 1.22 & {\color{black}1.34}          & \cdots& \cdots\\

 \text{} & \text{Total} & {\color{black}90.4} & 100    & \cdots& \cdots\\

 5^3 \text{P}_2& {\color{black}B^*D^*_2(2460)} & {\color{black}13.4}& {\color{black}18.8} & \cdots& \cdots\\
 & B^*D(2550) & 9.33 & {\color{black}13.3} & \cdots& \cdots\\
 \text{} & BD & 6.84 & {\color{black}9.32}      & \cdots& \cdots\\
 \text{} & B^*D & 6.55 &{\color{black}8.8}           & \cdots& \cdots\\
 \text{} & B^*D^* & 5.15 & {\color{black}7.62}       & \cdots& \cdots\\
 & {\color{black}B^*_2(5747)D} & {\color{black}5.4}& {\color{black}7.58} & \cdots& \cdots\\
 & {\color{black}BD_1(2430)} & {\color{black}4.8}& {\color{black}6.73} & \cdots& \cdots\\
 & {\color{black}BD^{\prime}_1(2420)} & {\color{black}4.47}& {\color{black}6.27} & \cdots& \cdots\\
 & {\color{black}B^*_2(5747)D^*} & {\color{black}4.1}& {\color{black}5.75} & \cdots& \cdots\\
 & {\color{black}B^*D_1(2430)} & {\color{black}3.17}& {\color{black}4.45} & \cdots& \cdots\\
 & {\color{black}BD^*_2(2460)} & {\color{black}2.42}& {\color{black}3.4} & \cdots& \cdots\\
 & {\color{black}B_1(5721)D} & {\color{black}1.56}& {\color{black}2.19} & \cdots& \cdots\\
 & {\color{black}B^*D^{\prime}_1(2420)} & {\color{black}1.22}& {\color{black}1.71} & \cdots& \cdots\\
  \text{} & {\color{black}B_s^*D_s^*} &{\color{black}1.11} & {\color{black}1.56}          & \cdots& \cdots\\

 \text{} & \text{Total} & {\color{black}71.3} & 100      & \cdots& \cdots\\
\bottomrule[1pt] \bottomrule[1pt]
\end{array}\]}
\end{table}

\subsection{$D$-wave states}
Partial widths and branching ratios of the OZI-allowed strong decay and radiative transition for $D$-wave $B_c$ states are shown in Tables \ref{d1}-\ref{d4}, Table \ref{dianfusheS} and Table \ref{2D2F}. 
For {\color{black} the} $B_c(1^3D_3)$ state, its main decay mode is $B_c(1^3D_3)\to B_c(1^3P_2)\gamma$, which has a branching fraction to be $\sim100\%$. So this state may decay via the radiative decay chain $B_c(1^3D_3)\to B_c(1^3P_2)\gamma\to B_c(1^3S_1)\gamma\gamma$.
Compared with the predicted partial width in Ref. \cite{Li:2019tbn}, the value of radiative transition $1D_2\to1P_1 \gamma$ is {\color{black}7} times smaller. Our calculated partial widths of the dominant radiative decay $1D_2\to1^3P_2 \gamma$ is roughly consistent with the value 8.7 keV given by Ref. \cite{Li:2019tbn}. 
\par

 In the strong decay case,  $BD^*$  is dominant decay  mode of $2D^\prime$ state, which have the width and branching ratio {\color{black} 138} MeV and {\color{black} 0.51}, respectively.
For the $2D$ state, $B^*D^*$  and $B^*D$  are its  main  decay  modes, which have the width of {\color{black} 45} MeV and {\color{black} 36} MeV.
 $B_c(3D_2^{\prime})$ are governed by the $B^*D$ mode with a branching fraction
\begin{eqnarray}
Br[B_c(3D^\prime_2)\to B^*D]\simeq 38\%.\nonumber
\end{eqnarray}

The $B_c(3^3D_1)$ and $B_c(3^3D_3)$ states all mainly decay into $B^*D^*$ channel, too.
The corresponding strong decay width are
\begin{eqnarray}\nonumber
&&\Gamma[B_c(3^3D_1)\to B^*D^*]\simeq {\color{black}14} ~\mathrm{MeV},\\\nonumber
&&\Gamma[B_c(3^3D_3)\to B^*D^*]\simeq 44 ~\mathrm{MeV}.\\\nonumber
\end{eqnarray}
Furthermore, the mass of $3D$ and $3D^\prime$ lie on the threshold of $B_s^*D_s^*$. And $B_sD_s^*$, $B_s^*D_s$, $B_s^*D_s^*$  channels have some contribution for the decays of $3D$ and $3D^\prime$. 
  It is found that the $B_c(3D_2^{\prime})$ state has a relatively narrow width of{\color{black} 65} MeV, which is half of that in Ref. \cite{Li:2019tbn}. 
  $4D$ and $4D^\prime$ have more strong decay channels for the higher mass and have the total width of about {\color{black} 133} MeV and {\color{black} 121} MeV, respectively. Branching ratios of main $B_c$ decay modes $BD^*$, $B^*D^*$ and $B^*D$ for $2^3D_1$ state are calculated to be $55 \%$, $22 \%$, and $16 \%$, respectively. One can see that $B^*D^*$ is the leading decay channel and the following $B^*D$ and $BD$ are the next-to-leading decay modes for the $B_c(2^3D_3)$. $B_c(3^3D_1)$ and $B_c(3^3D_3)$  dominantly decay to  $B^*D^*$ final channel, and $BD$ are their next-to-leading decay mode. 

The higher $4^3D_1$ and $4^3D_3$  states of $B_c$ mesons are also studied in present work and the corresponding strong decay widths are listed in Tables \ref{d1}-\ref{d4}. 
There are more strong decay channels for this two 4D-wave $B_c$ states. The total widths of  $4^3D_1$ and $4^3D_3$  are 130 and 184 MeV, respectively. 

\begin{table}
\caption{{\color{black}Partial widths and branching ratios of OZI-allowed strong decay for the mixture of the $2^1D_2$ and $2^3D_2$ states and $3^1D_2$ and $3^3D_2$ states. The width results are in units of MeV.}}
\vspace{-12pt}
\label{d1}
{\color{black}\[\begin{array}{clrrrr}
\toprule[1pt]\toprule[1pt]
\text{State}&\text{Channels}&\text{ This work}&\text{ Br ($\%$)}   &$ \cite{Li:2019tbn}$ &\text{ Br ($\%$)}\\
\midrule[0.7pt]
 2 \text{D}'_{2}& BD^*   &  {\color{black}138} &  {\color{black}50.6}          &66.8     &40.7\\
  & B^*D              &  {\color{black}76.5}  &  {\color{black}28.1}          &57.1     &34.7\\
                & B^*D^*            & {\color{black} 57.9}  &  {\color{black}21.2}          &40.4     &24.6\\
                & \text{Total}      &  {\color{black}272}   & 100           &164      &100\\
 
 2 \text{D}_{2} & B^*D^*            & {\color{black}45.3}  & {\color{black}47.2}          &12.3     &9\\             
                & B^*D              &{\color{black} 36.1}  & {\color{black}37.6}          &38.2     &27\\
                & BD^*              & {\color{black}14.6}  & {\color{black}15.2}          &89       &64\\
                & \text{Total}      & {\color{black}96.1}    & 100           &139      &100\\
            
3 \text{D}'_{2} & B^*D              & {\color{black} 24.6}  &  {\color{black}37.6}          &45.8     &34.6\\
                & B^*D^*            & 18.6  &  {\color{black}28.6}          &21.1     &16\\
                & BD^*              & 15.2  & {\color{black} 23.4}          &20.6     &15.6\\
                & B_{1}(5721)D      &  {\color{black}3.73}  &  {\color{black}5.71}          &\cdots   &\cdots\\
                & B^*_{s}D^*_{s}    & 1.4   & 2.11          &9.07     &6.8\\
                & B^*_{2}(5747)D    &  {\color{black}0.735} & 1.14          &\cdots   &\cdots\\
                & B_{s}D^*_{s}      &  {\color{black}0.619}  &  {\color{black}0.947}         &6.33   &4.8\\
                & B^*_{s}D_s        & {\color{black} 0.221} &  {\color{black}0.338}        &2.25     &1.7\\
                & BD^*_{0}(2300)    &\cdots &\cdots         &14.4     &10.9\\
                & B(1^3P_0)D        &\cdots &\cdots         &12.1     &9.1\\
                & \text{Total}      &  {\color{black}65.4}  & 100           &132      &100\\        
 
 3 \text{D}_{2} & B^{*}_{2}(5747)D  & 93.3  & 76.6          &\cdots   &\cdots\\
                & B^*D^*            & 18.9  & 15.5          &22.1     &19\\
                & B_{1}(5721)D      & {\color{black}2.77}  &{\color{black} 2.27}         &2.82     &2.5\\
                & BD^*              & {\color{black}2.16}  & {\color{black}1.77}          &13.8     &12\\
                & B^*_{s}D_s        & 2.07  & 1.7           &3.89     &3.4\\
                & B^*_{s}D^*_{s}    & 1.35  & 1.11          &11.6     &10\\
                & B_{s}D^*_{s}      & 1.14  & 0.936         &6.46     &5.7\\
                & B^*D              & 0.14  & 0.115         &38.9     &34.2\\ 
                & B^*D^*_{0}(2300)        &\cdots &\cdots         &13.6     &12\\
                & \text{Total}      & 122   & 100           &114      &100\\       
 \bottomrule[1pt] \bottomrule[1pt]
\end{array}\]}
\end{table}   

\begin{table}
\caption{{\color{black}Partial widths and branching ratios of OZI-allowed strong decay for the mixture of the $4^1D_2$ and $4^3D_2$ states. The width results are in units of MeV.}}
\vspace{-12pt}
\label{d2}
{\color{black}\[\begin{array}{clrrrr}
\toprule[1pt]\toprule[1pt]
\text{State}&\text{Channels}&\text{ This work}&\text{ Br ($\%$)}\\
\midrule[0.7pt]  
 4 \text{D}'_{2} & B^*_{2}\text{(5747)}D^* & {\color{black}27} &{\color{black}22.3}  \\
  \text{} & B^*D^*_{2}\text{(2460)} & {\color{black}21.7} & {\color{black}17.9} \\
   \text{} & B^*D & {\color{black}17.3} & 14.4 \\
   \text{} & B^*D^{'}_{1}(2420) & {\color{black}15.6} & {\color{black}12.9} \\
    \text{} & BD^*_{2}\text{(2460)} & {\color{black}12.1} & {\color{black}10} \\
 \text{} & B^*D_1\text{(2430)} & {\color{black}7.88} & {\color{black}6.51}\\
  \text{} & B^*_{2}\text{(5747)D} & {\color{black}5.05} &{\color{black}4.17} \\
   \text{} & B^*D^* & {\color{black}4.63} & {\color{black}3.83} \\
 \text{} & B_1\text{(5721)}D^* & {\color{black}2.95} & {\color{black}2.44} \\
 \text{} & B_1\text{(5721)D} & {\color{black}2.35} & {\color{black}1.94} \\
  & {\color{black}BD^*}    & {\color{black}1.85}& {\color{black}1.53}     \\
 
 \text{} & \text{Total} & {\color{black}121} & 100\\
 
 4 \text{D}_{2} & B^*D^*_{2}\text{(2460)} & 32.3 & 20.3 \\
 \text{} & B^*_{2}\text{(5747)}D^* & 23.7 &{\color{black} 17.7} \\
  \text{} & B^*_{2}\text{(5747)D} & {\color{black}22.9} & {\color{black}17.2} \\
   \text{} & BD^*_{2}\text{(2460)} & 18.3 & {\color{black}13.7} \\
    \text{} & B^*D_{1}^{\prime}(2420) & {\color{black}10.6} & {\color{black}7.95 }\\
 \text{} & B^*D_1\text{(2430)} & {\color{black}8.49} & {\color{black}6.37 }\\
  \text{} & B^*D^* & {\color{black}4.61} & {\color{black}3.46} \\
   \text{} & B_1\text{(5721)D} & {\color{black}2.21} & {\color{black}1.66} \\
    \text{} & B^*D & {\color{black}1.89} & {\color{black}1.42} \\
 \text{} & B_1\text{(5721)}D^* & {\color{black}1.25} & {\color{black}0.938} \\
 \text{} & \text{Total} & {\color{black}133} & 100 \\
\bottomrule[1pt] \bottomrule[1pt]
\end{array}\]}
\end{table}

 \begin{table}
\caption{{\color{black}Partial widths and branching ratios of OZI-allowed strong decay for $2^3D_1-3^3D_3$-wave $B_c$ states. The width results are in units of MeV.}}
\vspace{-12pt}
\label{d3}
{\color{black}\[\begin{array}{clrrrr}
\toprule[1pt]\toprule[1pt]
\text{State}    &\text{Channels}  &\text{ This work} &\text{ Br ($\%$)} &$ \cite{Li:2019tbn}$  &\text{ Br ($\%$)}\\
\midrule[0.7pt]
 2^3 \text{D}_1 & BD^*           & 47.5    & 54.7            &50.1   &87\\
                & B^*D^*         & 19      & 21.9            &0.48   &0.8\\
                & B^*D           & 13.5    & 15.6            &6.24   &10.9\\
                & BD             &{\color{black} 4.2 }   &{\color{black} 4.8}            &0.55   &1.0\\
                & B_{s}D_{s}     & 2.85    & 3.28            &0.18   &0.3\\
                & \text{Total}   & {\color{black}87.3}    & 100             &57     &100\\  

 2^3 \text{D}_3 & B^*D^*         & 140     & 61              &87     &46.1\\
                & B^*D           & 42.3    & 18.4            &50.8   &26.9\\
                & BD             & 39.5    & 17.2            &41.6   &22.1\\
                & BD^*           & 7.73    & 3.37            &9.29   &4.9\\
                & \text{Total}   & 230     & 100             &189    &100\\ 
                
3^3 \text{D}_1 & B^*D^*          & {\color{black}13.9 } & {\color{black}45.3}            &19.5   &21.9\\               
               & BD^*            & {\color{black}7.41 }   &{\color{black} 24.1}              &0.48   &0.5\\
               & B_{1}(5721)D    & {\color{black}3.46}    & {\color{black}11.3}            &0.76 &0.9\\
               & B_{s}^*D_{s}^*  & 2.24    &{\color{black} 7.27}            &16.5   &18.5\\
               & B_{s}D_{s}      & 1.29    & {\color{black}4.2}            &2.27   &2.5\\
               & B^*_{s}D_{s}    & {\color{black}0.797}   & {\color{black}2.6}            &3.16   &3.5\\
                 & BD              & {\color{black}0.782}    & {\color{black}2.6}            &25.2   &28.2\\
               & B_{2}^*(5747)D  & {\color{black}0.328}   &{\color{black} 1.07}            &\cdots &\cdots\\
               & B^*D            & {\color{black}0.28}   &{\color{black} 0.912}          &5.65   &6.3\\
               & B_{s}D^*_{s}    & {\color{black}0.142}   & {\color{black}0.463}           &1.82   &2.0\\
               & \text{Total}    & {\color{black}30.7}    & 100             &89     &100\\
               
3^3 \text{D}_3 & B^*D^*          &{\color{black} 43.9}    & {\color{black}55.9 }           &18.4   &16.4\\               
               & BD              & {\color{black}14.8}      & {\color{black}18.9}           &19.3   &17\\
               & B^*D            & {\color{black}14.2}    & {\color{black}18.1}            &29.7   &26.5\\
               & B_{1}(5721)D    & {\color{black}1.25}    & {\color{black}1.59}           &4.62   &4.1\\
               & BD^*            & {\color{black}1.64}    & {\color{black}2.09 }           &20.8   &18.6\\
               & B_{2}^*(5747)D  & 1.08    &{\color{black} 1.38}           &\cdots &\cdots\\
               & B_{s}^*D_{s}^*  & 0.824   &{\color{black} 1.05}           &6.6    &5.9\\
               & B_{s}D_{s}^*    & 0.628   & {\color{black}0.803}           &2.94   &2.6\\
               & B_{s}D{s}       &\cdots   &\cdots           &1.45   &1.3\\
               & B^{*}D_0^*(2300)&\cdots   &\cdots           &8.14   &7.3\\
               & \text{Total}    & 83.9    & 100             &112    &100\\
\bottomrule[1pt] \bottomrule[1pt]
\end{array}\]}
\end{table}

\begin{table}
\caption{{\color{black}Partial widths and branching ratios of OZI-allowed strong decay for $4^3D_1-4^3D_3$-wave $B_c$ states. The width results are in units of MeV.}}
\vspace{-12pt}
\label{d4}
{\color{black}\[\begin{array}{clrrrr}
\toprule[1pt]\toprule[1pt]
\text{State}&\text{Channels}&\text{ This work}&\text{ Br ($\%$)}\\
\midrule[0.7pt]
 4^3 \text{D}_1 \text{} & B^*_2\text{(5747)}D^* & 30.1& {\color{black}30.5 }\\
  \text{} & BD(2550) & 20 & {\color{black}20.3} \\
   \text{} & B^*D^*_{2}\text{(2460)} & 16.3 & {\color{black}16.4} \\
   \text{} & B^*_2\text{(5747)D} & {\color{black}6.81} & {\color{black}6.9} \\
 & B^*D_1\text{(2430)} & {\color{black}5.59 }& {\color{black}5.66} \\
  \text{} & B^*D_{1}^{\prime}(2420) & {\color{black}3.05} & {\color{black}3.09} \\
   \text{} & BD &{\color{black} 3.02} & {\color{black}3.06} \\
    \text{} & B^*D^* & {\color{black}2.94} &{\color{black} 2.98} \\
     \text{} & B_1\text{(5721)D} &{\color{black} 2.77} & {\color{black}2.8 }\\
      \text{} & BD^*_{2}\text{(2460)} &{\color{black} 1.86} &{\color{black} 1.88 }\\
       \text{} & {\color{black}BD^*} & {\color{black}1.16} & {\color{black}1.17} \\
        \text{} & B^*_sD^*_s &{\color{black} 1.12} &{\color{black}1.13} \\
         \text{} & BD_{1}^{\prime}(2420) &{\color{black} 1.09} &{\color{black} 1.1} \\
 \text{} & B_1\text{(5721)}D^* & {\color{black}1.06} & {\color{black}1.07 }\\
 \text{} & \text{Total} & {\color{black}98.8}& 100 \\
 
  4^3 \text{D}_3 & B^*D^*_{2}\text{(2460)} & 47.5 & {\color{black}30.6}\\
 \text{} & B^*_2\text{(5747)}D^* & 34.1 & {\color{black}22.2} \\
  \text{} & B^*D & 10.7 & {\color{black}6.77} \\
  \text{} & BD & {\color{black}9.86} & {\color{black}6.35} \\
 \text{} & B^*D^* & {\color{black}8.37} & {\color{black}5.39} \\
 \text{} & B^*D_1\text{(2430)} & {\color{black}7.8} & {\color{black}5.03} \\
  \text{} & BD^*_{2}\text{(2460)} & 6.87 & {\color{black}4.43} \\
 \text{} & BD_1\text{(2430)} & {\color{black}6.37} & {\color{black}4.11} \\
  \text{} & BD_{1}^{\prime}(2420) & {\color{black}5.58} & {\color{black}3.6} \\
   \text{} &{\color{black} B^*D_{1}^{\prime}(2420)} & {\color{black}4.81} & {\color{black}3.1} \\
   \text{} & B^*_2\text{(5747)D} & {\color{black}4.74} & {\color{black}3.05 }\\
    \text{} & BD^* & {\color{black}3.25} & {\color{black}2.09 }\\
     \text{} & B_1\text{(5721)}D^* & 1.61 & {\color{black}1.03} \\
 \text{} & B^*_sD^*_s & {\color{black}1.29} & {\color{black}0.831} \\
 \text{} & \text{Total} & {\color{black}155} &100 \\
\bottomrule[1pt] \bottomrule[1pt]
\end{array}\]}
\end{table}

\subsection{$F$-wave states}

{In the following, we will focus on the higher $nF$-wave $B_c$ states with $n=1,2,3$. The theoretical mass values of $F$-wave $B_c$ states are presented in
Table \ref{mass}.}
From our predictions of the decay properties for these $F$-wave states, it is found that the $B_c(1F_3^{\prime})$ state decay via radiative transitions $B_c(1F_3^{\prime})\to 1^3D_3 \gamma$, $B_c(1F_3^{\prime})\to 1D_2^{\prime} \gamma$ and $B_c(1F_3^{\prime})\to 1D_2 \gamma$. Our predicted partial width
\begin{eqnarray}\nonumber
\Gamma[B_c(1F_3^{\prime})\to 1^3D_3 \gamma]\simeq 4~\mathrm{keV},\\\nonumber
\Gamma[B_c(1F_3^{\prime})\to 1D_2^{\prime} \gamma]\simeq 1~\mathrm{keV},\\\nonumber
\Gamma[B_c(1F_3^{\prime})\to 1D_2 \gamma]\simeq {\color{black}73}~\mathrm{keV},\\\nonumber
\end{eqnarray}
our total radiative decay width  is very close to 82 keV predicted in Ref. \cite{2004Spectroscopy}. Other radiative transition properties can be found in Table \ref{dianfusheS} and Table \ref{2D2F}.
the $B_c(1F_3^{\prime})$ state also has strong decay $B_c(1F_3^{\prime})\to B^*D$, the predicted partial width
\begin{eqnarray}\nonumber
\Gamma[B_c(1F_3^{\prime})\to B^*D]\simeq {\color{black}0.426}~\mathrm{MeV}.\\\nonumber
\end{eqnarray}

The other $1F$ states $B_c(1F_3)$ has similar decay channels.
$B^*D^*$ is the leading decay channel of  $B_c(2F)$ and $B_c(2F^\prime)$ and the branching ratio is larger than  0.5.
$B^*D$ and $BD^*$  are the  next-to-leading decay channels of $B_c(2F)$ and $B_c(2F^\prime)$, respectively.
 The total widths of $B_c(2F)$ and $B_c(2F^\prime)$ are {\color{black} 127} MeV and 133 MeV, respectively. $B^*D^*$ has the similar contribution for the $B_c(3F)$ and $B_c(3F^\prime)$ states {\color{black}} and these two states have the total width of {\color{black} 135} MeV and {\color{black} 155} MeV.
These two states have different  leading decay channel, which are {\color{black} $B^*D^{\prime}_1(2420)$} and {\color{black} $BD^*_2(2460)$}.

$B_c(1^3F_4)$ only has two strong decay channels $BD$ and $B^*D$,  which contribute 78\% and 22\% for the total width. The results for the total width in this work and the results in Ref. \cite{Li:2019tbn} are all small, which are {\color{black} 0.99} MeV and 0.88 MeV, respectively.
Similar with $B_c(2F)$ and $B_c(2F^\prime)$, $B^*D^*$ is also the leading decay channel of  $B_c(2^3F_2)$ and $B_c(2^3F_4)$. The total width of this two states are 124 MeV and 153 MeV, respectively. 
$B_c(3^3F_2)$  dominantly decays to  $B^*D^*$ final channel, and {\color{black} $B^*D_1(2430)$} is its next-to-leading decay mode. 
{\color{black} $B^*_2(5747)D$} and {\color{black} $BD$} are the important decay channels. The total width for this state is about {\color{black} 109} MeV.
 For the $B_c(3^3F_4)$ state,  $B^*D$,  $B^*D^*$, $BD$ and  {\color{black} $B^*_2(5747)D^*$} have the  widths of 14 MeV, 12 MeV, 10 MeV, and {\color{black} 23} MeV, respectively, which are the main decay modes. Compared to Ref. \cite{Li:2019tbn}, our results have smaller decay width for the $3F_3$, $3F^{\prime}_2$ and $3^3F_4$ states

There are many strong decay channels for these higher mass $F$-wave states. Other detail of decay information can be seen in Tables \ref{f1}-\ref{f3}. 

\begin{table}[htbp]
\caption{{\color{black}Partial widths and branching ratios of OZI-allowed strong decay for the mixture of the $1^1F_3$ and $1^3F_3$ states $-$ $3^1F_3$ and $3^3F_3$ states. The width results are in units of MeV.}}
\vspace{-12pt}
\label{f1}
{\color{black}\[\begin{array}{clrrrr}
\toprule[1pt]\toprule[1pt]
\text{State}&\text{Channels}&\text{ This work}&\text{ Br ($\%$)}&$\cite{Li:2019tbn}$&\text{ Br ($\%$)}\\
\midrule[0.7pt]
 1 \text{F}'_{3} & B^*D & {\color{black}0.426} & 100         &15.1    &100\\
 \text{} & \text{Total} & {\color{black}0.426} &100 &15.1    &100\\
 
 1 \text{F}_{3} & B^*D & {\color{black}27.6} & 100  &8.53    &100\\
 \text{} & \text{Total} & {\color{black}27.6} & 100& 8.53    &100\\
 
 2 \text{F}'_{3} & B^*D^* & 71.4 & 53.6  & 80.3    &45\\
 \text{} & B^*D & 53.2 & 39.9  & 45.2    &25\\
 \text{} & BD^* & 5.7 & 4.28          &41.0    &23\\
 \text{} & B_sD^*_s & 1.98 & 1.49   &4.53    &3\\
 \text{} & B^*_sD_s & 1.04 & 0.78  &7.19    &4\\
 \text{} & \text{Total} & 133 & 100 &178     &100\\
 
 2 \text{F}_{3} & B^*D^* & 69.2 & 54.2    &90.2    &52\\ 
 \text{} & BD^* & 49.1 & 38.5       &30.2    &17  \\ 
 \text{} & B^*D & {\color{black}6.1} & {\color{black}4.79 }      &43.9    &25   \\
 \text{} & B^*_sD_s & 2.75 & 2.16   & 7.78& 4.5 \\
  & B_{s}D^{*}_{s}  & \cdots& \cdots &2.57    &1.5  \\
 \text{} & \text{Total} & {\color{black}127} & 100  &175     &100\\
 
  3 \text{F}'_{3}\text{} &{\color{black}BD^*_2(2460)}   &{\color{black}53.9}& {\color{black}34.8  }  &27.1 &8.9\\ 
  & B^*D & {\color{black}27.2} & {\color{black}17.6} & 33.6& 11\\
   \text{} &{\color{black} B^*(5747)D^*} &{\color{black}20.2}& {\color{black}13}& 36.7& 12 \\
     \text{} &{\color{black} B^*(5747)D} &{\color{black}18.2}& {\color{black}11.8}& 36.7& 12 \\
 \text{} & B^*D^* & {\color{black}16.1} & {\color{black}10.4} & 59.9& 19.6\\
  \text{} & {\color{black}B^*D_1(2430)} & {\color{black}6.62} & {\color{black}4.28} & \cdots& \cdots\\
   \text{} &B^*D_{1}^{\prime}(2420)    &{\color{black}4.59}& {\color{black}2.96}     &38.9 &12.8\\
    \text{} &{\color{black}BD_{1}(2430)}    &{\color{black}2.48}& {\color{black}1.6}     &38.9 &12.8\\
      \text{} &B_1\text{(5721)} D    &{\color{black}1.93} & {\color{black}1.25}     &\cdots &\cdots\\
 \text{} & B_sD^*_s & {\color{black}1} & {\color{black}0.646}& 1.69& 0.6 \\
 \text{} & {\color{black}B_s^*D^*_s} & {\color{black}0.961} & {\color{black}0.621}& {\color{black}4.85}& {\color{black}1.6} \\

  \text{} &B_1\text{(5721)} D^*    &{\color{black}0.794}& {\color{black}0.513}      &7.75 &2.5\\
 
 \text{} & BD^* &{\color{black}0.544}& {\color{black}0.351}& 34.4& 11.3 \\
 \text{} &B(1P^{'})D^*    &\cdots& \cdots       &30.4 &10\\
  & B_{s}D^*_{s0}(2317)  &\cdots   &\cdots  & 3.03  & 1.0 \\
  & B_{s}(1^3P_0)D_s      &\cdots   &\cdots  & 1.38  & 0.45 \\
 \text{} & \text{Total} & {\color{black}155} & 100& 305& 100 \\
 
 3 \text{F}_{3}  \text{} & B^*D_{1}^{\prime}(2420) & {\color{black}33.5}   &{\color{black}24.8}  &57.4  & 17.5 \\
    \text{} & B^*_2(5747)D & {\color{black}28.9}  &{\color{black}21.4} &27.6  & 8.4 \\
 & BD^* & {\color{black}16.9} & {\color{black}12.5}& 31.3& 9.5\\
  \text{} & B^*D_1(2430) & {\color{black}16.1}  &{\color{black}11.9} &\cdots  & \cdots\\ 
 \text{} & B^*D^* & {\color{black}14.7}& {\color{black}10.9}& 63.5& 19.3 \\
       \text{} & B_1\text{(5721)}D^* & {\color{black}7.79}   &{\color{black}5.77}  &2.3 & 0.7 \\
 \text{} & B^*D & {\color{black}6.51} & {\color{black}4.82} & 39.9& 12\\
     \text{} & BD_{1}(2430) & {\color{black}3.69}   &{\color{black}2.73}  &8.23 & 2.5 \\
            \text{} &B_1\text{(5721)}D & {\color{black}2.46}  &{\color{black}1.82} &2.26 & 0.7 \\
                \text{} & B^*_2(5747)D^* & {\color{black}1.75}  &{\color{black}1.3} &\cdots  & \cdots \\
  \text{} & {\color{black}BD^*_2(2460)} & {\color{black}1.37}   &{\color{black}1.01}  &34.2  & 10 \\
   \text{} & BD_{1}(2430) & \cdots   &\cdots  &16.6 & 5.1 \\
   \text{} & B(1P^{'})D^* & \cdots   &\cdots  &8.69  & 2.6 \\
    \text{} & B(1^3P_0)D^* & \cdots   &\cdots  &8.06  & 2.5 \\
      \text{} & B(1P^{'})D & \cdots   &\cdots  &6.25  & 1.9\\
      \text{} & B_s^*D_s^* & \cdots   &\cdots  &3.63  & 1.1\\
      \text{} & B_s^*D_s & \cdots   &\cdots  &2.64  & 0.8\\ 
      \text{} & B_sD_s^* & \cdots   &\cdots  &2.02  & 0.6\\ 
       \text{} & B^*D^*_{0}(2300)& \cdots   &\cdots  &1.73  & 0.53 \\
 \text{} & \text{Total} & {\color{black}135} & 100& 329& 100\\
  \bottomrule[1pt] \bottomrule[1pt]
\end{array}\]}
\end{table}

\begin{table}
\caption{{\color{black}Partial widths and branching ratios of OZI-allowed strong decay for $1^3F_2-3^3F_4$-wave $B_c$ states. The width results are in units of MeV.}}
\vspace{-12pt}
\label{f3}
{\color{black}\[\begin{array}{clrrrr}
\toprule[1pt]\toprule[1pt]
\text{State}&\text{Channels}&\text{ This work}&\text{ Br ($\%$)} &$\cite{Li:2019tbn}$&\text{ Br ($\%$)} \\
\midrule[0.7pt]
 1^3 \text{F}_2 & BD & 54.7 & 77.8 &61.9   &85\\
 \text{} & B^*D & 15.6 & 22.2             & 11.1   &15\\
 \text{} & \text{Total} & 70.3 & 100  &73     &100\\
 
 1^3 \text{F}_4 & BD & {\color{black}0.815} &{\color{black} 82.3} &0.85   &97\\
 \text{} & B^*D & {\color{black}0.112} & {\color{black}11.3}             & 0.03   &3\\
 \text{} & \text{Total} &{\color{black} 0.991} & 100  & 0.88   &100\\
 
 2^3 \text{F}_2 & B^*D^* & 102 & 82.3   &151    &68\\
 \text{} & BD & 13.8 & 11.1    &45.1   &20.2\\
 \text{} & B^*D &{\color{black} 2.83} &{\color{black} 2.29}    &19.2   &8.6\\
 \text{} & BD^* & {\color{black}2.77} & {\color{black}2.24}         &  0.39   &0.2 \\
 \text{} & B^*_sD_s & 1.11 & {\color{black}0.907}    &3.63&1.6\\
 \text{} & B_sD^*_s & 0.735 & 0.593    &3.17   &1.4\\
 \text{} & B_sD_s & 0.713 & 0.575         &0.68   &0.3\\
 \text{} & \text{Total} & 124 & 100& 223    &100\\
 
 2^3 \text{F}_4 & B^*D^* & 75.4 & 49.4     &57    &43\\
 \text{} & B^*D & 29.1 & 19                     &20.9  &16\\
 \text{} & BD^* & 28 & 18.3                    &37.7  &29\\
 \text{} & BD & 19.1 & 12.5            &8     &6\\
 \text{} & B_sD_s & 0.7 & 0.458           &4.48  &3.4\\
 \text{} & B^*_sD_s & 0.474 & 0.31    &3.26  &2.5\\
 \text{} & \text{Total} & 153 & 100     &131   &100\\
 
 3^3 \text{F}_2 & B^*D^* & {\color{black}27.5} &{\color{black}25.3}  &72 &31.5\\
 \text{} & B^*D_1(2430) & {\color{black}15.6} & {\color{black}14.3 }    &\cdots &\cdots\\
 &B_2^*(5747)D&{\color{black}10.7}& {\color{black}9.83}&12.1 &5.3\\
 \text{} & BD & {\color{black}10.3} & {\color{black}9.46}     &32.1 &14\\
 \text{} & BD^{\prime}_1(2420) & {\color{black}8.75} & {\color{black}8.04}     &\cdots &\cdots\\
 \text{} & B^*D^{\prime}_1(2420) & {\color{black}8.13} & {\color{black}7.47}     &\cdots &\cdots\\
 &BD_{1}(2430)  &{\color{black}6.76}& {\color{black}6.21}&2.3  &1\\
  &BD^*_{2}(2460)  &{\color{black}5.73}& {\color{black}5.26}&\cdots  &\cdots\\
  & B_{1}(5721)D &{\color{black}4.28}& {\color{black}3.93}&5.03 &2.2\\
    & B_{1}(5721)D^* &{\color{black}4}& {\color{black}3.67}&\cdots &\cdots\\
 \text{} & B^*D & {\color{black}3.33} & {\color{black}3.06}    &16.1 &7\\
  &B_2^*(5747)D^*&{\color{black}2.37}& {\color{black}2.18}&\cdots &\cdots\\
 \text{} & B^*_sD^*_s & 0.79 & {\color{black}0.73}   &5.25 &2.3\\
&B^{*}D_0^*(2300) &\cdots& \cdots&9.14 &4\\
  \text{} & \text{Total} & {\color{black}109} & 100       &228  &100\\
  
 3^3 \text{F}_4  \text{}  &B_2^*(5747)D^* &{\color{black}22.9}& {\color{black}23.8}  &\cdots &\cdots\\
 & B^*D & 14.4 & {\color{black}14.8}    &8.9 &5.0\\
 \text{} & B^*D^* & 12.2 & {\color{black}12.8}    &50.9 &28.7  \\
 \text{} & BD & 10.7 &{\color{black}11.1}  &2.82& 1.6          \\
 \text{}  &B_2^*(5747)D &{\color{black}10.6}& {\color{black}11}  &12.3 &6.9\\
 \text{} & BD^* & 10.4 & {\color{black}10.6} &20.2& 11.4    \\
 \text{}  &BD_{1}^{\prime}(2420)  &{\color{black}8.36}& {\color{black}8.67}   &9.19 &5.2\\
 &B_1\text{(5721)}D   &{\color{black}2.28}&{\color{black} 2.37}  &1.85 &1.04\\
 \text{} & B^*_sD^*_s & 1.5 & {\color{black}1.54} &9.19 &5.2   \\
 &B^{*}D_{1}^{\prime}(2420) &{\color{black}1.07}& {\color{black}1.11}      &2.83 &1.6\\
 &B_1\text{(5721)}D^{*}    &{\color{black}0.628}& {\color{black}0.651}        &3.03 &1.7\\
 \text{}  &BD_{1}(2430) &{\color{black}0.528}& {\color{black}0.548}  &11.2 &6.3\\
 \text{}  &{\color{black}BD^*_{2}(2460)} &{\color{black}0.496}& {\color{black}0.514} &11.2 &6.3\\
 \text{} & B_sD_s & 0.257 & {\color{black}0.26}     &3.2& 1.8       \\
 \text{} & B^*_sD_s & 0.136 & {\color{black}0.135}&3.2& 1.8     \\
 &B^{*}D_{1}(2430)&{\color{black}0.123}& {\color{black}0.128}  &8.42 &4.7\\
  \text{} & BD^* &\cdots& \cdots& 34.4& 11.3 \\
 \text{} &B(1P^{'})D    &\cdots& \cdots       &19.4 &10.9\\
&B^{*}D_0^*(2300) &\cdots& \cdots     &1.91 &1.1\\
 \text{} & \text{Total} & {\color{black}96.4} &100 &177  &100\\
\bottomrule[1pt] \bottomrule[1pt]
\end{array}\]}
\end{table} 

\subsection{$G$-wave states}
In this subsection, we sweepingly discuss the decay features of $G$-wave $B_c$ states. 

The decay properties are presented in Table \ref{g1} and Table\ref{GR}. It is found that all $G$-wave $B_c$ states decay via radiative transition and strong transition.
\par
The $B_c(1G^\prime)$ mainly decays into $BD^*$, $B^*D^*$ and $B^*D$ with the branching ratios 59$\%$, 24$\%$ and 18\% respectively. The corresponding partner $B_c(1G)$ dominantly decay to $B^*D$, which has the width and  branching ratio {\color{black} 88.3} MeV and 0.87. 

For the $B_c(2G)$ and $B_c(2G^\prime)$, $B^*D^*$ has the {\color{black} similar} width {\color{black}},  $BD^*$ and  $B^*D$ are their two important decay channels.
The total widths of these two states are 156 MeV and 67 MeV, respectively.

We predict that the main decay channels of $B_c(1^3G_3)$ are $BD$, $B^*D$ and $BD^*$ with corresponding
partial widths 64 MeV, 43 MeV, and 20 MeV, respectively. 
For the $B_c(1^3G_5)$ state, $BD$ and $B^*D^*$ are
important decay modes. It is found that this  state has a relatively narrow width of $\sim 45$ MeV.
Furthermore, the radiative transitions have negligible contribution for all these states. 

The $B_c(2^3G_3)$ dominantly decays into $B^*D^*$, and $BD$ channel is the next-to-leading decay channel. $B_c(2^3G_5)$ dominantly decays into $B^*D^*$, $BD^*$ and $B^*D$ channels. It has the total width of 54 MeV.

 We give our predictions of the $E1$ transitions and strong decays of $B_c(nG)$ ($n=1,2$). Combining these $E1$ transitions with their strong decays, we found that the branching fractions of the $E1$ transitions  are so small which can be be ignored.

\newpage
\renewcommand{\arraystretch}{1.5}
{\textcolor{black}{
\setlength{\tabcolsep}{1.5pt}
\begin{longtable*}{clccccccccccc}
\caption{Partial widths of the E1 dominant radiative transitions for the $2S$-, $1P$-, $1D$-, and $1F$-wave $B_c$ states.}
    \label{dianfusheS}\\
 \toprule[1pt]\toprule[1pt]
  \multicolumn{1}{c}{\multirow{1}*{~ }}&\multicolumn{1}{c}{\multirow{1}*{~}}& \multicolumn{5}{c} {$E_{\gamma}$  (MeV)}   &   \multicolumn{5}{c}{$\Gamma_{E1}$(keV)} 
  \\
   \cline{3-6} \cline{7-11}
 \multicolumn{1}{c}{\multirow{1}*{Initial state }}&\multicolumn{1}{c}{\multirow{1}*{ Final state}}&Ours&Ref. \cite{Li:2019tbn}&Ref. \cite{Gershtein:1994dxw} & Ref. \cite{2004Spectroscopy}&Ours&Ref. \cite{Li:2019tbn}&Ref. \cite{Gershtein:1994dxw}&Ref. \cite{2004Spectroscopy}\\
         \bottomrule[1pt] 
                             \endfirsthead
\multicolumn{12}{l}{Table \ref{dianfusheS} continued}\\
               \bottomrule[1pt] \bottomrule[1pt]
        \multicolumn{1}{c}{\multirow{2}*{~}}& \multicolumn{1}{c}{\multirow{2}*{~}}&  \multicolumn{5}{c}{$E_{\gamma}$  (MeV)}   &   \multicolumn{5}{c}{$\Gamma_{E1}$ (keV)}  \\
         \cline{3-6} \cline{7-11}
  \multicolumn{1}{c}{Initial state} &\multicolumn{1}{c}{Final state}&Ours &Ref. \cite{Li:2019tbn}&Ref. \cite{Gershtein:1994dxw} & Ref. \cite{2004Spectroscopy}&Ours&Ref. \cite{Li:2019tbn}&Ref. \cite{Gershtein:1994dxw}&Ref. \cite{2004Spectroscopy}\\
               \bottomrule[1pt] 
                \endhead
               \bottomrule[1pt] 
           \bottomrule[1pt]
            \endlastfoot
$2^1S_0$ &$1P^{'}_{1}$  &110.1  & 96    &138  &104   & 7.56   &6.38  &15.9  &6.1 \\
         &$1P_{1}$      &101.2    &114   &150  &113   &{\color{black}1.2}   &5.33 &1.9   &1.3\\
$2^3S_1$ &$1^3P_{2}$    &122.8    &102       &159  &118   &7.28   &6.98   &14.8  &5.7\\
         &$1P^{'}_{1}$  &144.4    &115   &173  &136   &{\color{black}1.03}   &1.56  &1.0   &0.7\\
         &$1P_{1}$      &135.6    &133      &185  &144   &4.2    &4.62   &12.8  &4.7\\
         &$1^3P_{0}$    &192.2      &174     &219  &179 &3.45   &3.48    &7.7   &2.9\\
$1^3P_2$ &$1^3S_{1}$    &445.3      &445      &426  &416 &90.04  &87     &102.9 &83\\
$2^3P_2$ &$1^3D_{3}$    &131.7    &129    &127  &118   &9.78   &14     &10.9  &6.8\\
         &$1D^{'}_{2}$  &127.8    &127     &118  &122   &{\color{black}0.66}    &0.93   &0.5   &0.6\\
         &$1D_{2}$      &136.7    &135     &133  &127   &1.02   &1.1    &1.5   &0.7\\
         &$1^3D_{1}$    &136.7    &139      &126  &135   &0.112  &0.13    &0.1   &0.1\\
         &$2^3S_{1}$    &263      &265      &232  &272   &49.37  &50     &49.4  &55\\
         &$1^3S_{1}$    &802.3    &785     &817  &778   &30.35  &52     &25.8  &14\\
$1P^{'}_1$ &$1^3S_1$    &424.7    &433       &412  &399   &{\color{black}13.75}  &40    &8.1   &11\\
           &$1^1S_0$    &469.5      &484     &476  &462   &76.5   &74     &131.1 &80\\
$2P^{'}_1$ &$1D^{'}_{2}$    &109    &117    &108 &113     &0.96     &1.05    &3.5    &5.5\\
           &$1D_{2}$        &118    &125   &123 &123     &{\color{black}7.0}     &0.03     &2.5    &1.3\\
           &$1^3D_{1}$      &118    &129   &116 &121     &{\color{black}0.47}      &1.27     &0.3    &0.2\\
           &$2^3S_{1}$      &244.7  &255   &222 &258     &8.8      &25       &5.9    &5.5\\
           &$1^3S_{1}$      &785    &777   &807 &769     &{\color{black}4.17}      &26      &2.5    &0.6\\
           &$2^1S_{0}$      &278    &274   &257 &289     &43       &36        &58.0   &52\\
           &$1^1S_{0}$      &827.6  &825     &871 &825     &24.42    &44       &131.1  &19\\
$1P_1$     &$1^3S_1$        &433    &416    &400 &391   &71.52     &70      &77.8 &60\\
           &$1^1S_0$        &477.8  &468    &464 &454   &{\color{black}16.5}     &35    &11.6 &13\\
$2P_1$     &$1D^{'}_{2}$    &117      &101    &97 &108     &{\color{black}6.4}      &0.006   &1.2  &0.8\\
           &$1D_{2}$        &125.8    &109   &112 &103    &0.748    &0.84      &3.9  &3.6\\
           &$1^3D_{1}$      &125.8    &113    &105 &116    &1.94     &1.45     &1.6  &1.6\\
           &$2^3S_{1}$      &252.4    &240    &211 &253    &38.48    &34.0     &32.1 &45.0\\
           &$1^3S_{1}$      &792.5    &762    &796 &761    &14.67    &40      &15.3 &5.4\\
           &$2^1S_{0}$      &286      &258    &246 &284    &{\color{black}13.62}    &19      &8.1  &5.7\\
           &$1^1S_{0}$      &834.7    &811   &860 &820    &{\color{black}7.32}      &25     &3.1  &2.1\\
$1^3P_0$   &$1^3S_{1}$      &378.6    &377  &366 &358    &59.58    &96    &65.3 &55\\
$2^3P_0$   &$1^3D_{1}$      &80.5     &86    &80  &93     &3.24     &5.6    &3.2  &4.2\\
           &$2^3S_{1}$      &208      &214    &186 &231    &33.49    &53        &25.5 &42\\
           &$1^3S_{1}$      &751.5    &738     &771 &741    &3.19     &41       &16.1 &1\\
$1^3D_3$   &$1^3P_{2}$      &254      &239       &264 &272    &65.98    &67       &76.9 &78\\
$1D^{'}_{2}$ &$1^3P_{2}$    &258      &241      &273 &263    &7.57    &8.3       &6.8  &8.8\\
             &$1P^{'}_{1}$  &279      &253      &287 &280    &{\color{black}16.3}   &41        &46   &63\\
             &$1P_{1}$      &271      &271     &301 &289    &{\color{black}50.45}   &0.39       &25   &7\\
$1D_{2}$     &$1^3P_{2}$    &249.4    &233     &258 &268    &8.71     &8.7        &12.2  &9.6\\
             &$1P^{'}_{1}$  &270.6    &246      &272 &285    &{\color{black}56}    &1.09       &18.4  &15\\
             &$1P_{1}$      &262      &263      &284 &294    &{\color{black}6.28}  &44        &44.6  &64\\
$1^3D_1$   &$1^3P_{2}$      &249.4    &229      &265 &255    &1.74   &0.7      &2.2   &1.8\\
           &$1P^{'}_{1}$    &270.6    &242      &279 &273    &{\color{black}5.12}   &12         &3.3   &4.4\\
           &$1P_{1}$        &262      &259       &291 &281    &22.79  &29        &39.2  &28\\
           &$1^3P_{0}$      &317.5    &299      &325 &315    &52.36  &65         &79.9  &55\\
$1^3F_4$   &$1^3D_{3}$      &204      &194      &-   &222    &63.93   &69         &-     &81\\
$1F^{'}_3$ &$1^3D_{3}$      &215.7    &207      &-   &227    &3.59    &4.76       &-     &5.4\\
           &$1D^{'}_{2}$    &211.8    &205      &-   &231    &1.32    &32        &-     &82\\
           &$1D_{2}$        &220.5    &212      &-   &236    &{\color{black}73.2}   &0.04     &-     &0.04\\
$1F_3$     &$1^3D_{3}$      &203    &191      &-   &218    &4.0     &4.91      &-     &3.7\\
           &$1D^{'}_{2}$    &199.2  &189     &-   &222    &56.24   &0.22      &-     &0.5\\
           &$1D_{2}$        &208    &197      &-   &226    &{\color{black}0.58}    &29        &-     &78\\
$1^3F_2$   &$1^3D_{3}$      &212.8    &202     &-   &221    &0.32   &0.12      &-     &0.4\\
           &$1D^{'}_{2}$    &208.9    &200     &-   &224    &5.0    &5.72     &-     &5.3\\
           &$1D_{2}$        &217.6    &208     &-   &229    &6.57   &6.36      &-     &6.5\\
           &$1^3D_{1}$      &217.6    &212     &-   &221    &61.39  &78        &-     &75\\
        \bottomrule
\end{longtable*}}}

\begin{table*}[htp]
\caption{{\color{black}{Partial widths and branching ratios of OZI-allowed strong decay for $G$-wave $B_c$ states. The width results are in units of MeV.}}}
\vspace{-12pt}
\label{g1}
{\color{black}
\[\begin{array}{clccclcc}
\toprule[1pt]\toprule[1pt]
\text{State}&\text{Channels}&\text{ This work}&\text{ Br ($\%$)}&\text{State}&\text{Channels}&\text{ This work}&\text{ Br ($\%$)}\\
\midrule[0.7pt]
 1 \text{G}'_{4} & BD^* & {\color{black}45.5} & 58.5 &1^3 \text{G}_3 & BD & 63.7 & 46.6 \\
 \text{} & B^*D^* & {\color{black}18.4} & 23.6 &\text{} & B^*D & 42.5 & 31.1\\
 \text{} & B^*D & {\color{black}13.8} & 17.9 &\text{} & BD^* & 20.3 & 14.9 \\
 &&&&\text{} & B^*D^* & 9 & 6.59\\
 &&&& \text{} & B_sD_s & 0.916 & 0.671 \\
 \text{} & \text{Total} & {\color{black}77.8}& 100& \text{} & \text{Total} & 137 & 100 \\
 1 \text{G}_{4} & B^*D & {\color{black}88.3} &{\color{black} 87}&  1^3 \text{G}_5 & B^*D^* & 28.6 & 64.1 \\
 \text{} & B^*D^* & 11.6 & 11.2 & \text{} & BD & 9.04 & 20.3 \\
 \text{} & BD^* & {\color{black}1.64} &{\color{black} 1.62 }& \text{} & B^*D & 6.18 & 13.8\\
 \text{} & B^*_sD_s & 0.126 & 0.122& \text{} & BD^* & 0.821 & 1.84 \\
 \text{} & \text{Total} & {\color{black}102} & 100& \text{} & \text{Total} & 44.6 & 100\\
 2 \text{G}'_{4} & B^*D^* &{\color{black} 41} & {\color{black}61.5}  & 2^3 \text{G}_3 & B^*D^* & 74.1 & 57.5\\
 \text{} & B^*D & {\color{black}12.3 }& {\color{black}18.5}& \text{}                    & BD & 23.6 & 18.9 \\
 \text{} & BD^* & {\color{black}7.73} & {\color{black}11.6} & \text{}                   & B^*D & 13.1 & 10.5\\
 \text{} & B^*_sD^*_s & 1.49 & 2.23     &  \text{}       & B^*_2(5747)D & 10.6 & 8.28\\
 \text{} & B_1(5721)D & {\color{black}1.31} & {\color{black}1.97 }& \text{}    & {\color{black}B_1 (5721)D} & {\color{black}7.62} & {\color{black}5.69}\\ 
 \text{} & B^*_sD_s & 1.17 & 1.75 & \text{}               & BD^* & 3.3 & 2.58 \\
 \text{} & B_sD^*_s& 0.88 & 1.32 & \text{}                & B^*_sD^*_s & 1.2 & 0.937\\
 \text{} & B{}^*_2(5747)D & 0.693 & 1.04 & \text{}        & B_sD^*_s & 0.384 & 0.3 \\
 \text{} & \text{Total} & 66.8 & 100 & \text{}            & \text{Total} & {\color{black}134} & 100\\
 2 \text{G}_{4} & B{}^*_2 (5747)D & 56.8 & 36.5 & 2^3 \text{G}_5 & B^*D^* & 22.4 & 41.1 \\
 \text{} & B^*D^* & 40.1 & 25.8 & \text{} & BD^* & 15.4 & 28.3\\
 \text{} & B^*D & 28.5 & 18.3 &\text{} & B^*D & {\color{black}7.56} & {\color{black}14.1} \\
 \text{} & BD^* & 27.8 & 17.9 & \text{} & BD & {\color{black}3.12} & {\color{black}5.82} \\
 \text{} & B^*_sD^*_s & 1.32 & 0.848&  \text{} & B^*_sD^*_s  & 2.55 & 4.68 \\
 \text{} & B_1(5721)D & 0.561 & 0.36 & \text{} & B_1(5721)D & 0.699 & 1.28\\
 \text{} & B_sD^*_s & 0.551 & 0.354 & \text{} & B^*_2(5747)D & {\color{black}0.237} &{\color{black} 0.442} \\
 \text{} & \text{Total} & 156 & 100&\text{} & \text{Total} & 54.4 & 100\\
\bottomrule[1pt] \bottomrule[1pt]
\end{array}\]}
\end{table*}

\renewcommand{\arraystretch}{1.2} 
\setlength{\tabcolsep}{2.5mm}{
\begin{table*}[htbp]
		\centering
	\caption{{\color{black}	Partial widths of the E1 dominant radiative transitions for the $2D$- and $2F$-wave $B_c$ states.}}
		\label{2D2F}
{\color{black}\begin{tabular}{cccccccc}
			\hline
				\hline
Initial state & Final state & $\Gamma_{E1}$ (keV) & Initial state & Final state& $\Gamma_{E1}$ (keV) \\
\hline
$2^3D_1$  & $1P^{'}_1$ &{\color{black}0.526} &$2^3D_3$   &$1^3P_{2}$  &{\color{black}7.46}\\
          & $1P_1$    &{\color{black}2.47}      &           &$2^3P_{2}$   &{\color{black}48.7}\\
          & $2P^{'}_1$ &{\color{black}5.12}      &           &$1^3F_{4}$  &{\color{black}7.28}  \\ 
          & $2P_1$  &{\color{black}15.5}      &           &            &\\
          & $1^3P_0$   &{\color{black}13.5}      &           &              & \\
          & $1^3P_{2}$ &{\color{black}0.1}       &           &             & \\
          & $2^3P_0$     &{\color{black}37.2}      &           &             &\\ 
          & $2^3P_2$    &{\color{black}1.29}      &           &          & \\ 
          & $1^3F_{2}$  &{\color{black}5.43}      &           &         & \\ 
$2D^{'}_2$  & $1P^{'}_1$ &{\color{black}1.7}       &$2D_2$     &$2P^{'}_1$   &{\color{black}50.2}\\
            & $1P_1$     &{\color{black}7.36}      &           &$2P_1$     &{\color{black}2.6}\\
            & $2P^{'}_1$  &{\color{black}8.4}       &           &$2^3P_{2}$   &{\color{black}6.36}  \\ 
            & $2P_1$  &{\color{black}39.54}     &           &$1F_{3}$     &{\color{black}0.05}\\
            & $1^3P_{2}$  &{\color{black}0.6}       &           &         &  \\
            & $2^3P_2$  &{\color{black}5.51}      &           &             &  \\ 
            & $1F_{3}^{'}$  &  {\color{black}0.11}    &           &          &\\ 
$2^3F_2$  &$2D^{'}_{2}$    &{\color{black}3.83}     &$2^3F_4$       &$1^3D_{3}$   &{\color{black}4.96} \\  
          &$2D_{2}$    &{\color{black}4.54}     &               &$2^3D_{3}$   &{\color{black}51.2} \\
          &$1^3D_{1}$  &{\color{black}6.3}      &               &$1^3G_{5}$  &{\color{black}5.38} \\
          &$2^3D_{1}$  &{\color{black}48.9}     &               &            & \\ 
          &$1^3G_{3}$  &{\color{black}4.61}     &               &             & \\ 

 $2F_3^{'}$   &  $2D_2^{'}$    &{\color{black}1.05}     &$2F_3$         &$1D_{2}$  &{\color{black}0.06}     \\ 
              &  $1D_2$        &{\color{black}5.73}     &               &$2D_{2}$  &{\color{black}0.29}\\
              &  $1G_4$    &{\color{black}7.29}     &               &$2^3D_{3}$  &{\color{black}3.12}      \\
              &  $1G^{'}_{4}$     &{\color{black}0.1}      &                  &        &  \\
              &  $1D_2^{'}$    &{\color{black}0.11}     &               &      &        \\
                \hline
                 \hline
\end{tabular}}
\end{table*}}

\renewcommand{\arraystretch}{1.2} 
\setlength{\tabcolsep}{2.5mm}{
\begin{table*}[htbp]
		\centering
	\caption{{\color{black}	Partial widths of the E1 dominant radiative transitions for the $G$-wave $B_c$ states.}}
		\label{GR}
{\color{black}\begin{tabular}{cccccccccc}
			\hline	\hline
Initial state & Final state & $\Gamma_{E1}$ (keV) & Initial state & Final state  & $\Gamma_{E1}$ (keV) \\
\hline
$1^3G_3$  &$1F^{'}_{3}$    &{\color{black}2.25}     &$2^3G_3$   &$1F^{'}_{3}$  &{\color{black}0.145}\\
          &$1^3F_{2}$       &{\color{black}62.4}     &           &$2F^{'}_{3}$   &{\color{black}1.88}  \\ 
          &$1F_{3}$      &{\color{black}3.68}     &           &$1^3F_{2}$    &{\color{black}4.55}  \\
          &$1^3F_{4}$      &{\color{black}0.1}      &           &$1F_{3}$     &{\color{black}0.21} \\
          &               &         &           &$1^3F_{4}$  &{\color{black}0.006}\\ 
          &             &         &           &$2^3F_{2}$   &{\color{black}52.86} \\ 
          &            &         &           &$2F_{3}$   &{\color{black}3.05} \\ 
          &           &         &           &$2^3F_{4}$    &{\color{black}0.083} \\ 
          &              &         &           &$1^3H_{4}$  &{\color{black}3.8} \\
$1^3G_5$  &$1^3F_{4}$    &{\color{black}64.7}     &$2^3G_5$   &$1^3F_{4}$    &{\color{black}3.92}\\ 
          &              &         &           &$2^3F_{4}$     &{\color{black}52.59
          }\\
&          &               &           &$1H^{'}_{5}$  &{\color{black}0.042} \\
          &            &         &           &$1^3H_{4}$  &{\color{black}0.0008} \\
          &            &         &           &$1H_{5}$      &{\color{black}0.08}   \\
$1G^{'}_{4}$  &$1F^{'}_{3}$   &{\color{black}1.43}  &$1G_{4}$  &$1F^{'}_{3}$  &{\color{black}48.83} \\  
              &$1F_{3}$           &{\color{black}78.66}      &               &$1F_{3}$     &{\color{black}0.83} \\
              &$1^3F_{4}$       &{\color{black}2.2}        &               &$1^3F_{4}$      &{\color{black}2.23} \\
$2G^{'}_{4}$  &$1F^{'}_{3}$   &{\color{black}0.08}     &$2G_{4}$       &$1F^{'}_{3}$  &{\color{black}3.78}  \\ 
              &$2F^{'}_{3}$   &{\color{black}1.1}      &               &$2F^{'}_{3}$  &{\color{black}40.97}  \\
              &$1F_{3}$      &{\color{black}4.48}     &               &$1F_{3}$      &{\color{black}0.059} \\
              &$1^3F_{4}$    &{\color{black}0.123}    &               &$1^3F_{4}$    &{\color{black}0.145} \\
              &$2F_{3}$      &{\color{black}64.9}     &               &$2F_{3}$     &{\color{black}0.639} \\
              &$2^3F_{4}$     &{\color{black}1.82}     &               &$2^3F_{4}$   &{\color{black}1.84}  \\
              &$1H^{'}_{5}$    &{\color{black}0.057}    &               &$1H^{'}_{5}$  &{\color{black}2.36} \\
              &$1^3H_{4}$     &{\color{black}0.07}    &               &$1^3H_{4}$    &{\color{black}0.056} \\
              &$1H_{5}$       &{\color{black}6.23}     &               &$1H_{5}$     &{\color{black}0.081} \\
          \hline	\hline
\end{tabular}}
\end{table*}}

\section{conclusion}\label{sec5}
  The mass spectra of $B_c$ mesons are studied in this paper using Cornell potential with the screening effect. The parameters of the potential model by fitting $B_c$, $b\bar{b}$, $c\bar{c}$, $B$, $B_s$, $D$ and $D_s$ mesons are given, we also predict the masses of the excited states of $B_c$ mesons.   
According to our research, the mass of the higher exited states in this work is lower than most other potential models, which is caused by the screening effect. 
For the radiative decays of $B_c$ mesons, we obtain the decay widths of the E1 dipolefor their $S-G$ waves. The decay widths are all small, within $1$ MeV. We also study the decay width of the S-wave $B_c$ meson state to G-wave $B_c$ meson state using the $^3P_0$ model and give the branching ratios. 
Analysis the strong two-body decay of the $B_c$ mesons, the important decay channel is $B^*D^*$ for many $B_c$ mesons. In addition, 
$B^*_sD_s$ and $B_s D_s^*$ make small contribution. Some main results are emphasized as follows.

For the $S$-wave states, the low-lying $S$-wave states $B_c^*(2S)$ and $B_c(2S)$ only decay via radiative transitions. Other $S$-wave states all decay via radiative transitions and strong decay and $B^*D^*$ makes large contribution for almost all higher $S$-wave states. For the $P$-wave states, it is found that the decay modes of the $1P$ only has $E1$ decay processes with small total width less than $10^{-1}$ MeV and $BD$ channel is the most important decay mode of $B_c(2^3P_2)$ state. For the $D$-wave and $F$-wave states, some of them present some radiative decay channels with large widths which  are allowed to observation. This may help for establishing the $D$-wave and $F$- wave $B_c$ states in the future experiments.  All the $G$-wave $B_c$ states have strong decays and radiative transitions, but the contribution of radiative transitions is very small.

We expect that our research can provide some helpful information for searching for $B_c$ mesons in future experiments.

\section{ACKNOWLEDGMENTS}
We are very grateful for professor Xiang Liu's help. This work is supported by the Science and Technology Department of Qinghai Province No. 2019-ZJ-A10 and National Natural Science Foundation of China under Grants No. 11965016 and 12247101. T-Y. L. and L. T. contributed equally to this work.

\bibliographystyle{apsrev4-1}
\bibliography{ref}
\end{document}